\documentclass[10pt,aps,prd,nofootinbib,twocolumn,superscriptaddress,floatfix,notitlepage]{revtex4-2}

\usepackage[dvipsnames]{xcolor}
\usepackage{graphicx}
\usepackage{amsmath,amsfonts,amssymb}
\usepackage[breaklinks,colorlinks,urlcolor=Plum,citecolor=MidnightBlue,linkcolor=WildStrawberry]{hyperref}
\usepackage{enumitem}
\usepackage{orcidlink}
\usepackage{array}
\usepackage{tabularx}
\usepackage{bbm}

\newcommand{\td}[0]{\mathrm{d}}
\newcommand{\eq}[0]{\vec{\mathbbm{e}}_q}
\newcommand{\ep}[0]{\vec{\mathbbm{e}}_p}

\newcommand{\fvec}[1]{\vec{\mathbbm{f}}_{\rm {#1}}}
\newcommand{\avec}[1]{\vec{\mathbbm{a}}_{\rm {#1}}}
\newcommand{\svec}[1]{\vec{\mathbbm{s}}_{\rm {#1}}}
\newcommand{\vvec}[0]{\vec{\mathbbm{v}}}
\newcommand{\Tvec}[1]{\vec{\mathbb{T}}_{\rm {#1}}}
\newcommand{\sdiff}[0]{\vec{\mathbbm{s}}_{\rm {e}}^{~\rm (diff)}}
\newcommand{\sbs}[0]{\vec{\mathbbm{s}}_{\rm {i}}^{~\rm (bs)}}
\newcommand{\Lmat}[1]{\mathbb{L}}

\newcommand{\papertitle}{Optomechanical transfer factors for scattered light noise estimations in the beamtubes of ground-based gravitational wave detectors}

\begin{document}

\title[]{\papertitle}

\author{M.~Andr\'es-Carcasona\orcidlink{0000-0002-8738-1672}}
\email{mandresc@mit.edu (corresponding author)}
\affiliation{LIGO Laboratory, Massachusetts Institute of Technology, Cambridge, MA 02139, USA}
\affiliation{Kavli Institute for Astrophysics and Space Research, Massachusetts Institute of Technology, Cambridge, MA 02139, USA}


\begin{abstract}
Scattered light is a relevant noise source in current ground-based gravitational-wave detectors and a critical design concern for next-generation observatories. Beamtube scattered light estimates usually combine optical propagation simulations with analytical couplings that do not fully propagate the frequency-dependent optomechanical response of the interferometer to the strain readout. In this work, we use improved analytical transfer factors to convert scattered-light field perturbations in the Fabry--Pérot arm cavities into equivalent strain noise, consistently including radiation-pressure coupling, signal-extraction dynamics, microscopic detunings, and the homodyne readout angle. The formalism keeps the full amplitude and phase quadrature content of the scattered field, including the cross terms that arise in both diffraction and backscattering noise. For diffraction, we also identify the regime in which the linearized coupling to baffle motion is valid, avoiding unnecessary phase wrapping. Using representative LIGO, Cosmic Explorer (CE), and Einstein Telescope - Low Frequency (ET-LF) configurations, we show that legacy estimates are recovered in phase dominated regimes, but can differ when radiation-pressure coupling, quadrature correlations, or detuned signal extraction become important. In particular, the revised ET-LF estimate changes substantially with respect to previous beamtube noise budgets due to the detuned signal extraction cavity. These results provide a more complete framework for scattered light noise estimations for present and future gravitational-wave detectors.
\end{abstract}

\maketitle

\section{Introduction}

Gravitational-wave (GW) detectors are precision instruments whose scientific outcome strongly depends on maintaining a low level of noise across the observation band. Thanks to the latest instrumental improvements in the ground based network of detectors LIGO-Virgo-KAGRA (LVK)~\cite{VIRGO:2014yos,LIGOScientific:2014pky,Capote:2024rmo,Aso:2013eba}, hundreds of events have been detected so far~\cite{LIGOScientific:2018mvr,LIGOScientific:2020ibl,LIGOScientific:2021usb,KAGRA:2021vkt,LIGOScientific:2025slb,LIGOScientific:2026wfs}. There are many noise mechanisms that can limit the detector sensitivity and we focus in this work on the one known as stray, or scattered, light. This noise originates from unwanted optical fields that re-couple to the main beam, which can imprint phase and amplitude fluctuations on the field, degrading the strain sensitivity and producing non-Gaussian transient artifacts (glitches) in the data~\cite{Accadia:2010zzb,Fiori:2020arj,Was:2020ziy,LIGO:2020zwl,Longo:2020onu,Longo:2021avq,Longo:2023vac,Bianchi:2021unp}.

Scattered-light noise is not only a known concern in current detectors~\cite{LIGO:2020zwl}, but also a critical design and commissioning risk for next-generation ground-based observatories such as the Einstein Telescope (ET)~\cite{ETcds,ETdesign} and Cosmic Explorer (CE)~\cite{CE1,CE2,CE3}. In these facilities, longer optical paths, more demanding sensitivity targets, and increasingly complex optical layouts tighten the requirements on stray-light control and on the fidelity of the associated noise modeling employed for its design. Reliable tools are essential both to guide mitigation strategies, such as baffle design and placement, and to interpret measured couplings during commissioning and operation.

Current beamtube scattered-light noise estimates for next-generation detectors typically combine FFT-based optical propagation tools, such as the Stationary Interferometer Simulation (SIS) code~\cite{Romero21}, with analytical expressions that recast the simulated optical perturbation into an equivalent strain-noise contribution~\cite{Andres-Carcasona:2023qom,Andres-Carcasona:2025xwq}. These expressions were developed for the initial design of LIGO and Virgo~\cite{Thorne89,FlanaganThorne95_Diff,Flanagan94,Flanagan95,Vinet96,Vinet97,Brisson98}, and later refined~\cite{Ottaway:2012oce,Hiro_ETMripple,CE_backscattering,Hall:2020dps}. While this approach has been highly valuable, a key limitation is that it generally does not include a realistic, frequency dependent transfer function between the additional field generated by scattered light and the final measured strain. As a result, the mapping from optical perturbation to detector output can miss the filtering and optomechanical resonances of the interferometer configuration used for GW readout.

In this paper, we derive the transfer functions required to convert a simulated (or measured) small perturbation of the main Fabry--Pérot (FP) arm resonant field into an equivalent strain noise. Our derivation follows the style and conventions of Refs.~\cite{McCuller:2021mbn,Hall:2024vsc,Miao:2018pai,Buonanno:2001cj,Danilishin:2012fa,Harms:2003hn} and is based on the two-photon formalism, enabling direct consistency with quantum-noise and interferometer-response treatments. Different optical design parameters (such as the reflectivity of the different mirrors, detuning of the cavities or readout schemes) can alter the final scattered light noise measured and it is important to account for it to accurately evaluate the final strain noise caused by scattered light in the main beamtubes.  

The optomechanical transfer functions used in this work are closely related to the standard input-output relations in quantum noise. In that context, the frequency dependent response of FP cavities, signal extraction cavities, radiation pressure, and homodyne readout has been extensively derived and used to predict quantum shot noise, radiation pressure noise, ponderomotive squeezing, and the effect of different readout quadratures \cite{Kimble:2000gu,Buonanno:2001cj,Danilishin:2012fa}. Our goal is therefore not to introduce a new framework, but rather to apply this known optomechanical response as a set of transfer factors that map classical scattered-light field perturbations inside the arm cavities to the final strain readout.

Realistic interferometer transfer functions have also been mentioned and used previously in the context of scattered light studies~\cite{Polini:2026wls}. In particular, Ref.~\cite{Was:2020ziy} derived and used detailed models of transfer factors for Advanced Virgo's end benches. The problem addressed here is nonetheless complementary. We focus on scattered-light perturbations generated within the main FP arm cavities, at the beamtube or baffle level. The corresponding source locations and field perturbations are not the same as those treated in Ref.~\cite{Was:2020ziy}. This motivates deriving the specific transfer factors needed to connect beamtube scattering simulations to equivalent strain noise and to do so for a more generic detector configuration.

\section{Model and conventions}
\begin{figure*}[htbp]
    \centering
    \includegraphics[width=\linewidth]{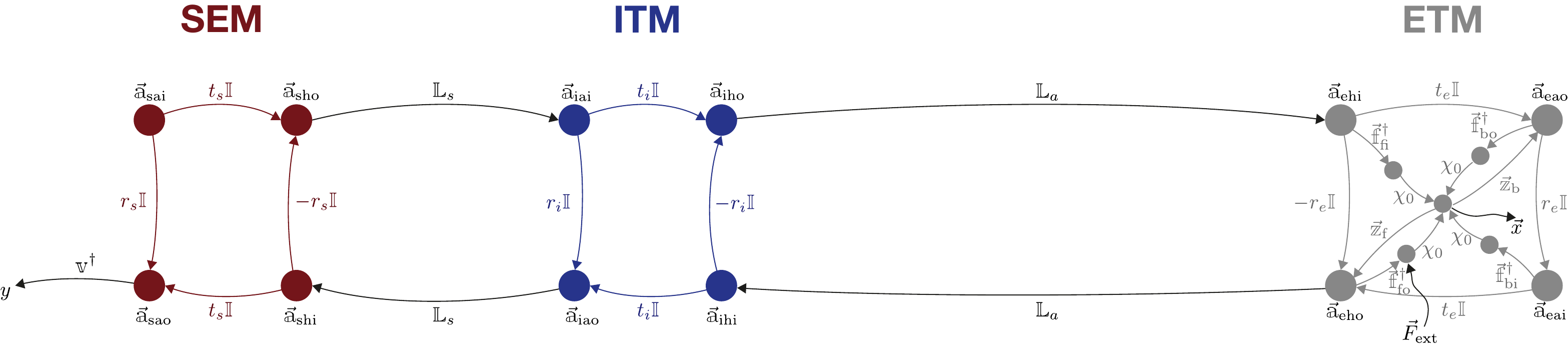}
    \caption{Simplified signal-flow diagram of the three-mirror coupled cavity used in this work, formed by the SEM, ITM, and ETM. The diagram indicates the propagation operators along the signal extraction (index $s$) and arm (index $a$) sections, and highlights the ETM optomechanical coupling via the mechanical susceptibility.}
    \label{fig:ccav_sfg}
\end{figure*}

We adopt an equivalent optical architecture consisting of a three mirror coupled cavity formed by the signal extraction mirror (SEM), the input test mass (ITM), and the end test mass (ETM). To a good approximation, the dynamics of the differential arm motion of a dual-recycled FP Michelson interferometer like ground based GW detectors can be described by this three mirror coupled cavity~\cite{Hall:2024vsc,Kuns:2026szt}. Figure~\ref{fig:ccav_sfg} summarizes the corresponding signal-flow diagram and the field relations used throughout the calculation.

To describe the optical field we use the notation of Ref.~\cite{McCuller:2021mbn}, where the state of the field is represented by the amplitude ($\hat{q}$) and phase ($\hat{p}$) quadratures as~\cite{Caves:1985zz,Schumaker:1985zz,Kuns:2026szt}

\begin{equation}
\vec{\mathbbm{a}}=
\begin{bmatrix}
\hat{q}\\
\hat{p}
\end{bmatrix}=\hat{q}~\vec{\mathbbm{e}}_q+\hat{p}~\vec{\mathbbm{e}}_p,
\qquad
\vec{\mathbbm{e}}_q=\begin{bmatrix}
1\\
0
\end{bmatrix},
\qquad
\vec{\mathbbm{e}}_p=\begin{bmatrix}
0\\
1
\end{bmatrix}.
\label{eq:2photon_def}
\end{equation}

In general, these quantities are functions of the frequency $\Omega = 2\pi f$. We denote each of the states by a set of three indices as $\avec{jkl}$ where $\rm j\in\{s,i,e\}$ denotes the mirror to which is referred, $\rm k\in\{h, a\}$  to denote either the high- (HR) or anti-reflective (AR) side of the mirror and $\rm l\in\{i,o\}$ to denote if the field is incoming or outgoing.

Each of these states can undergo some transformations corresponding to some physical interactions.  The propagation over a distance $L_i$ alters the quadratures as

\begin{equation}
\avec{out} = \mathbb{L}_i(\Omega)\avec{in}, , \quad \mathbb{L}_i(\Omega)= e^{-i \Omega L_i/c} \mathbb{R}(\varphi_i)\, ,
\label{eq:prop_def}
\end{equation}
where $\varphi_i$ is a microscopic detuning phase, and
\begin{equation}
\mathbb{R}(\phi)=
\begin{pmatrix}
\cos\phi & -\sin\phi\\
\sin\phi & \cos\phi
\end{pmatrix}\, ,
\label{eq:rot_def}
\end{equation}
is the quadrature rotation associated with that microscopic phase.

Each of the mirrors is described by the amplitude reflectivity and transmissivity denoted by $r$ and $t$, respectively. If the mirror has losses, $r^2+t^2< 1$. In the two-photon formalism, the reflection and transmission operators are computed as these scalar coefficients and the identity matrix (i.e. $r\mathbb{I}$ and $t\mathbb{I}$, respectively). We denote each mirror by the indices $s$, $i$ and $e$ to denote the SEM, ITM and ETM mirrors, respectively.

The signal flow diagram of Fig.~\ref{fig:ccav_sfg} explicitly denotes the effect of radiation pressure on the ETM. The vectors that translate the mirror motion to the excitation of optical fields in the phase quadrature are~(see Appendix A of Ref.~\cite{Hall:2024vsc})
\begin{equation}
    \vec{\mathbbm{z}} = \vec{\mathbbm{z}}_{\rm f} = 2kr\sqrt{P}~\ep\, ,
\end{equation}
where $k=2\pi/\lambda$, $\lambda$ is the laser wavelength and $P$ the laser power. Similarly, the radiation pressure forces are computed as~\cite{Hall:2024vsc}
\begin{equation}
    \fvec{}^\dagger=\fvec{fi}^\dagger = \frac{2}{c}\sqrt{P}~\eq^{\dagger},\, \fvec{fo}^\dagger = -r\fvec{}^\dagger,\, \fvec{bo}^\dagger = -t\fvec{}^\dagger,\, \fvec{bi}^\dagger=0\, ,
\end{equation}

Finally, after the field leaves the left-side of the SEM, it goes towards the detection port, where the readout takes place by beating the field outgoing from the AR side of the SEM, $\avec{sao}$, with a local oscillator, $\vvec$, such that the final measurement becomes
\begin{equation}\label{eq:readout}
    y = \vvec^\dagger \avec{sao}\, ,\qquad \vvec = \begin{bmatrix}
        \sin\zeta\\
        \cos\zeta
    \end{bmatrix}\, .
\end{equation}

Here, $\zeta$ denotes the homodyne angle, which represents the quadrature angle selected by the readout. In the two-photon formalism, the outgoing field $\avec{sao}$ is written in terms of amplitude and phase quadratures, and the readout projects this two-dimensional quadrature vector onto the direction $\vvec$. Keeping $\zeta$ as a free parameter is important because the coupling of scattered-light perturbations to the measured channel generally depends on the readout quadrature. In the presence of a cavity detuning, signal extraction, and radiation pressure, amplitude and phase perturbations are mixed by the interferometer response, so the measured scattered-light contribution can change significantly with $\zeta$.

This generic notation also allows the same transfer-function derivation to be applied to the most relevant readout implementations currently used (or considered) in GW interferometers. In DC readout, which is the standard scheme in advanced interferometers, a small differential arm (DARM) offset is introduced so that a controlled amount of carrier light leaks to the antisymmetric port and acts as an effective local oscillator. In this case, the measurement can still be described by Eq.~\eqref{eq:readout}, with $\zeta$ representing the effective homodyne angle set by the interferometer operating point and output optics. In balanced homodyne readout, an external local oscillator is interfered with the antisymmetric port field, and the local oscillator phase directly determines $\zeta$. This scheme is naturally described by the same projection formalism and is particularly convenient when discussing arbitrary quadrature measurements. By keeping $\zeta$ explicit in the derivations below, the resulting transfer factors remain directly applicable to a broad class of realistic readout configurations without modifying the optical model.

We summarize the parameters and notation employed throughout this work in Tab.~\ref{tab:variables}.

\begin{table}[htbp]
    \centering
    \begin{tabular}{c|l}
       \hline
       \textbf{Variable} & \textbf{Description} \\
       \hline\hline
       $r_e,\ t_e$ & Reflectivity/transmissivity of the ETM \\
       \hline
       $r_i,\ t_i$ & Reflectivity/transmissivity of the ITM \\
       \hline
       $r_s,\ t_s$ & Reflectivity/transmissivity of the SEM \\
       \hline
       $\varphi_a$ & Detuning phase  of the arm cavity \\
       \hline
       $\varphi_s$ & Detuning phase of the signal extraction cavity  \\
       \hline
        $L_a$ & Length of the main arm \\
       \hline
        $L_s$ & Length of the signal extraction cavity \\
       \hline
       $\omega$ & Laser angular frequency; $k\equiv\omega/c=2\pi/\lambda$ \\
       \hline
       $\lambda$ & Laser wavelength \\
       \hline
       $P$ & Circulating power in the FP cavity \\
       \hline
       $\zeta$ & Readout angle \\
       \hline
       $\chi_0$ & Test mass mechanical susceptibility \\
       \hline
    \end{tabular}
    \caption{Summary of notation used in this work. Parameters are treated as real scalars. Any effective optical loss may be absorbed into $r_k,t_k$ (with $r_k^2+t_k^2< 1$).}
    \label{tab:variables}
\end{table}

\section{Transfer functions} \label{sec:TFs}
The goal is to find the transfer factors between a perturbation incoming into the HR side of the ETM to the readout and the transfer factor from a perturbation outgoing from the HR side of the ITM to the readout, as these are the perturbations relevant to backscattering and diffraction noises.

If we denote a perturbation outgoing from the HR-ITM plane as $\svec{iho}(\Omega)$, we aim at finding a transfer factor of the form 
\begin{equation}
    y = \Tvec{i}^\dagger(\Omega)\svec{iho}(\Omega)\, ,
\end{equation}
where $y$ denotes the final readout and $\Tvec{i}^\dagger(\Omega)$ the transfer factor.

On the other hand, if we denote a perturbation of the field incoming to the HR side of the ETM as $\svec{ehi}(\Omega)$, then we are interested in the transfer factor to the final readout as 
\begin{equation}
    y = \Tvec{e}^\dagger(\Omega)\svec{ehi}(\Omega)\, .
\end{equation}

The first step to obtain these transfer factors is to reduce the ETM part of the flow diagram of Fig.~\ref{fig:ccav_sfg} to a mirror with effective reflectivity of $\mathbb{R}_{\rm e}$. We can ignore the transmissivity because we do not keep track of the field that exits the AR side of the ETM. The derivation of effective reflectivity is quite standard (see for example Appendix A of Ref.~\cite{Hall:2024vsc}), so we can directly use the result. This is~\cite{Kuns:2026szt} 
\begin{equation}
    \mathbb{R}_{\rm e}(\Omega) = -r_e\begin{bmatrix}
        1 & 0 \\
        -\mathcal{K}(\Omega) & 1
    \end{bmatrix}\\ ,
\end{equation}
where\footnote{The signal-flow diagram includes the radiation pressure response of the ETM explicitly. To account also for the ITM we take $\chi_0\to2\chi_0$, which is valid for equal ITM and ETM masses, differential arm mode and no suspension asymmetry.}
\begin{equation}
    \mathcal{K}(\Omega) = \frac{8\pi P \chi_0(\Omega) }{\lambda c}\left( 1+r_e^2-t_e^2\right)\, .
\end{equation}

This expression explicitly shows that when the effect of a back-reaction caused by the radiation pressure on the mirror is accounted for, the reflection operator is no longer diagonal. This implies that there is some cross-coupling of amplitude modulations into phase noise, proportional to $\mathcal{K}(\Omega)$. Such effect must be accounted for in order to realistically translate the scattered light field into noise. 

The mechanical susceptibility of the mirror, $\chi_0$, can modify this cross-coupling. In the case of a single free-mass mirror of mass $m$, the susceptibility is simply 
\begin{equation}\label{eq:free_chi}
    \chi_0(\Omega) = -\frac{1}{m\Omega^2}\, .
\end{equation}

For a more complex susceptibility model, we can include the information of resonances. For example a single resonance would yield a susceptibility of the form
\begin{equation}
    \chi_0(\Omega) = -\frac{1}{m}\frac{1}{\Omega^2-\Omega_n^2-i\frac{\Omega_n}{Q_n}\Omega}\, ,
\end{equation}
where $\Omega_n$ is the angular resonant frequency and $Q_n$ the $Q$-factor of that resonance.

With these terms, we can now fully derive the transfer factors. The derivation is rather standard and we present it in Appendix~\ref{app:derivation} for completeness. The final results are 
\begin{align}
    \Tvec{i}^\dagger(\Omega) & = \frac{t_st_i}{\sqrt{2}}  \vvec^\dagger [\mathbb{I}+r_s\mathbb{L}_{\rm s}\mathbb{R}_{\rm a}\mathbb{L}_{\rm s}]^{-1}\mathbb{L}_s[\mathbb{I}+r_i\mathbb{E}_{\rm a}]^{-1}\mathbb{E}_a\, ,\label{eq:Ti}\\
    \Tvec{e}^\dagger(\Omega) &= \frac{t_st_i}{\sqrt{2}}  \vvec^\dagger [\mathbb{I}+r_s\mathbb{L}_{\rm s}\mathbb{R}_{\rm a}\mathbb{L}_{\rm s}]^{-1}\mathbb{L}_s[\mathbb{I}+r_i\mathbb{E}_{\rm a}]^{-1}\mathbb{L}_a\mathbb{R}_e\, .\label{eq:Te}
\end{align}
where the effective arm reflectivity and the arm round-trip matrices are
\begin{align}
\mathbb{R}_{\rm a}&=r_i\mathbb{I}+t_i^2\left[\mathbb{I}+r_i\mathbb{E}_a\right]^{-1}\mathbb{E}_a\, ,\\ 
\mathbb{E}_{\rm a} &= \mathbb{L}_a\mathbb{R}_{\rm e}(\Omega)\mathbb{L}_a\, .
\end{align}

\subsection{Approximate expressions}\label{sec:approximate}
While Eqs.~\eqref{eq:Ti} and \eqref{eq:Te} represent the full transfer functions required, we can find some approximate expressions under some reasonable assumptions. If we assume that the propagation operators are $\mathbb{L}_j = e^{-i\varphi_j}\mathbb{I}$ with $\varphi_j=(\Omega-\delta\omega_j)L_j/c$, then we can simplify the transfer factors to 
\begin{align}
    & \Tvec{e}^\dagger=\frac{-r_e}{\sqrt{2}}t_{\rm cc}\vvec^\dagger [\mathbb{I}-K_{\rm eff}\mathbb{N}] \label{eq:approx_Ti}\,,\\
    & \Tvec{i}^\dagger=\frac{-r_ee^{-i\varphi_a}}{\sqrt{2}}t_{\rm cc}\vvec^\dagger [\mathbb{I}-K_{\rm eff}\mathbb{N}] \, ,
\end{align}
where
\begin{equation}
    \mathbb{N} = \begin{bmatrix}
        0 & 0\\ 
        1 & 0
    \end{bmatrix}\, ,
\end{equation}
and
\begin{equation}
    K_{\rm eff} = \frac{\mathcal{K}}{1-r_ir_ee^{-2i\varphi_a}}+\frac{r_sr_et_a^2e^{-2i\varphi_s}\mathcal{K}}{1+r_sr_ae^{-2i\varphi_s}}\, .
\end{equation}

We have also introduced the coupled-cavity transmission  $t_{\rm cc}$, the arm transmission $t_a$ and the arm reflectivity $r_a$. These can be approximated using the zero, pole, gain approximation in the Signal Recycling (SR) or Resonant Sideband Extraction (RSE) cases as~\cite{Kuns:2026szt}
\begin{equation}
    t_{\rm cc}(\Omega) = \begin{cases}
        \displaystyle\sqrt{\frac{4\mathcal{F}_a\mathcal{F}_s}{\pi^2}}\frac{1}{1+i(\Omega-\delta \omega_a)/\gamma_{\rm sr}},\qquad \mathrm{SR}\\ \\
        \displaystyle-\sqrt{\frac{\mathcal{F}_a}{\mathcal{F}_s}}\frac{i}{1+i(\Omega-\delta \omega_a)/\gamma_{\rm rse}},\qquad \mathrm{RSE}
    \end{cases}\, ,
\end{equation}
\begin{equation}
    r_a(\Omega) = -\frac{1-i(\Omega-\delta \omega_a)/\gamma_a}{1+i(\Omega-\delta \omega_a)/\gamma_a}\, ,
\end{equation}
\begin{equation}
    t_a(\Omega) = \sqrt{\frac{(1+r_i)\mathcal{F}_a}{\pi}}\frac{1}{1+i(\Omega-\delta\omega_a)/\gamma_a}
\end{equation}
where the finesses and poles are~\cite{Kuns:2026szt}
\begin{align}
    \mathcal{F}_a  = \frac{\pi}{1-r_i}&,\quad \gamma_a=\frac{\pi c}{2\mathcal{F}_aL_a}\, , \\
    \mathcal{F}_s = \frac{\pi}{1-r_s}&,\quad \gamma_s=\frac{\pi c}{2\mathcal{F}_sL_s}\, , \\
     \gamma_{\rm sr} = \frac{1-r_s}{1+r_s}\gamma_a\, &,\quad \gamma_{\rm rse} = \frac{1+r_s}{1-r_s}\gamma_a\, .
\end{align}

We can also simplify $K_{\rm eff}$ using the same approximations. In this case, we obtain 
\begin{widetext}
    \begin{equation} \label{eq:Keff_full}
    K_{\rm eff}(\Omega)=\begin{cases}
        \displaystyle \mathcal{K}(\Omega)\,
    \frac{\mathcal F_a}{\pi}\,
    \frac{1}{1+i(\Omega-\delta\omega_a)/\gamma_a}
    \left[
    1+
    \frac{r_s r_e(1+r_i)}{1-r_s}
    \frac{1}{1+i(\Omega-\delta\omega_a)/\gamma_{\rm sr}}
    \right]
    ,\qquad \mathrm{SR}\\ \\
        \displaystyle \mathcal{K}(\Omega)\frac{\mathcal F_a}{\pi}\,
\frac{1}{1+i(\Omega-\delta\omega_a)/\gamma_a}
\left[
1-
\frac{r_s r_e(1+r_i)}{1+r_s}
\frac{1}{1+i(\Omega-\delta\omega_a)/\gamma_{\rm rse}}
\right] ,\qquad \mathrm{RSE}
    \end{cases}\, ,
\end{equation}
\end{widetext}

This term becomes most relevant in the regime where radiation pressure dominates over phase noise. The frequency that determines the transition is the standard quantum limit (SQL) frequency, which for a free mass is~\cite{Kuns:2026szt}
\begin{equation}
    \Omega^{\rm fm}_{\rm SQL} = \sqrt{\frac{8k P}{m c}}\, .
\end{equation}
Then, if $\Omega^{\rm fm}_{\rm SQL}\ll \gamma_a,\gamma_{\rm rse}$ or $\Omega^{\rm fm}_{\rm SQL}\ll \gamma_a,\gamma_{\rm sr}$, we can simplify Eq.~\eqref{eq:Keff_full} to 
\begin{equation}\label{eq:Keff_red}
    K_{\rm eff}(\Omega)\simeq \mathcal{K}(\Omega)\Gamma^{\rm SR/RSE}\, ,
\end{equation}
with 
\begin{align}
    \Gamma^{\rm SR} &= \frac{1+r_ir_s}{1-r_i-r_s+r_ir_s}\, ,\\
    \Gamma^{\rm RSE} &= \frac{1-r_ir_s}{1-r_i+r_s-r_ir_s}\, .
\end{align}

Finally, for the typical conditions of operation (except for ET-LF) of RSE mode, tuned, phase readout, and for frequencies $\Omega\ll \gamma_{\rm rse}$ where stray light is mostly relevant, the transfer factors can be approximated up to an irrelevant global phase as 
\begin{align}
    & \Tvec{e}^\dagger=\frac{-r_e}{\sqrt{2}}\sqrt{\frac{\mathcal{F}_a}{\mathcal{F}_s}}[-\mathcal{K}\Gamma^{\rm RSE}\quad 1] \label{eq:approx_Te_final}\,,\\
    & \Tvec{i}^\dagger=\frac{-r_ee^{-i\varphi_a}}{\sqrt{2}}\sqrt{\frac{\mathcal{F}_a}{\mathcal{F}_s}}[-\mathcal{K}\Gamma^{\rm RSE}\quad 1]  \, .
\end{align}

\section{Noise estimations}
With the transfer functions derived in the previous section, we can now derive the strain noise power spectral densities (PSD) which can be included in the noise budget for GW detectors. 

The two dominant sources of scattered light noise in the main arms of ground-based detectors are that of backscattering and diffraction~\cite{Flanagan94,Thorne89}. Backscattering refers to the process where the light that has scattered from the optics interacts with a non-optical component, imprinting some modulation due to its vibration, and this can re-enter the main beam and recombine with it. When that happens, the additional modulation of the field can mimic a real GW, thereby degrading the sensitivity. Any surface exposed to light is a potential source of backscattered light noise and to minimize it, current detectors implement baffles, beam dumps and shrouds to absorb or redirect as much of this unwanted light as possible. Nonetheless, it is impossible to get rid of all of it, as even these pieces will be sources of such noise. Therefore, a correct modeling is necessary to ensure that they are engineered to ensure that this noise is not a limiting factor. 

Diffraction, on the other hand, is mainly caused by the clipping of the field by the finite inner apertures of the baffles of the main arms. When the beam passes through one of these baffles the inner edge clips it and this imprints a spatial structure. At each baffle the field gets clipped and this effect can be coherent, so that when the field gets to the other mirror, constructive patterns caused by diffraction at each edge can interfere with the circulating beam. The more baffles that are present in the cavity, the stronger this effect gets. This has been shown to be extremely problematic for the longer arms of next-generation ground-based detectors~\cite{Andres-Carcasona:2023qom,Andres-Carcasona:2025xwq}.

In the following subsections we derive the specific strain noise PSDs for backscattering and diffraction with the input that would be obtained by some model expansion~\cite{Vajente:22,VajenteCEbaffles,Andres-Carcasona:2026prv} or an FFT code, such as SIS~\cite{Romero21}. The latter is considered as the current standard in the design of future detectors.

\subsection{Diffraction}
Inside the FP cavity, the field can be expressed as a superposition of transverse electromagnetic (TEM) modes. The cavity is mainly designed to have the TEM$_{00}$ mode resonating. FFT-based codes can compute the resonating field inside the FP cavity and from it obtain the TEM$_{00}$ component, $\psi_{00}$. At the same time, it can compute the resonating field after this has been perturbed by a displacement of a baffle in the transverse plane (which is assumed to be the $x-y$ plane), and obtain as well its TEM$_{00}$ component, $\psi_{00}^{\rm (diff)}$. To do this, SIS takes the field emitted from the ITM, propagates it to the baffle which is displaced in the transverse plane by a small amount $\delta$, clips it, propagates it to the ETM and finally recombines it with the main beam~\cite{Romero21}. The overlap coefficient among the two fields is  
\begin{equation}
    a_{00} = \langle \psi_{00},\psi^{\rm (diff)}_{00}\rangle\, ,
\end{equation}
where $\langle a,b\rangle=\iint_{A} ab^*~\td x \td y$, denotes the inner product. In practice, SIS computes this coefficient for a baffle displaced by an amount $+\delta$ in the direction of the axis under consideration (typically either the $x$ or $y$ axis), and then by $-\delta$ so that it can estimate the first derivative of the coupling factor as\footnote{SIS in particular computes this coupling after the reflection at the ETM. We treat nonetheless the perturbation at the incoming node of the ETM so that the proper radiation pressure effect is accounted for. We show in Appendix~\ref{app:diffraction_node} that for the typical mirrors of GW detectors this error is negligible. This conclusion is also supported by the results presented in Sec.~\ref{sec:Results}.}  
\begin{equation}
    \delta a_{00} = \frac{a_{00}(+\delta)-a_{00}(-\delta)}{2\delta }\, .
\end{equation}

One can obtain the time domain perturbation of the TEM$_{00}$ component of the circulating field by multiplying this quantity by the motion of the scatterer, such that 
\begin{equation}
    \psi^{\rm (diff)}_{00}(t) = \psi_{00} [1+\delta a_{00}x_\perp(t)]\, ,
\end{equation}
where $x_\perp(t)$ is the vibration of the baffle in the transverse plane. This calculation is assuming that the motion is sufficiently small so that the linear coupling holds. We explain in Appendix~\ref{app:phase_wrapping_diffraction} the regime of validity of this assumption and the appearance of non-linearities, similar to phase-wrapping, when the assumption is not valid.

In general, this coupling will be a complex number and will generate some sidebands as 
\begin{equation}
    \sdiff = \begin{bmatrix}
        \delta a_{00}\tilde{x}_\perp(\Omega)\\
        \delta a_{00}^*\tilde{x}^*_\perp(-\Omega)
    \end{bmatrix}\, .
\end{equation}
For a real displacement of the baffle, $\tilde{x}^*_\perp(-\Omega)=\tilde{x}_\perp(\Omega)$. In the two photon picture this perturbation becomes
\begin{equation}
   \sdiff = \sqrt{2}\begin{bmatrix}
        \delta a_R\\
        \delta a_I
    \end{bmatrix}\tilde{x}_{\perp}(\Omega)\, ,
\end{equation}
where $\delta a_R$ and $\delta a_I$ are the real and imaginary parts of $\delta a_{00}$, respectively, and where we have used the common relation to convert between sideband and two-photon quadratures~\cite{Kimble:2000gu}
\begin{equation}
    \begin{bmatrix}
        \avec{q}(\Omega)\\
        \avec{p}(\Omega)
    \end{bmatrix}=\frac{1}{\sqrt{2}}\begin{bmatrix}
        1 & 1\\
        -i & i
    \end{bmatrix} \begin{bmatrix}
        \avec{u}(+\Omega)\\
        \avec{l}^*(-\Omega)
    \end{bmatrix}=\mathbb{A}\begin{bmatrix}
        \avec{u}(+\Omega)\\
        \avec{l}^*(-\Omega)
    \end{bmatrix}\, .
\end{equation}
The indices $u$ and $l$ denote the upper and lower sidebands, respectively.

The final readout becomes
\begin{equation}\label{eq:sediff}
    y^{\rm (diff)} = \Tvec{e}^\dagger \sdiff\, .
\end{equation}

To understand how this perturbation will affect the strain noise, we need to quantify what would be the effect of an arm length change $\delta L$ on the final measurement as this is the same physical effect that a real GW would have. The field in that case acquires a small phase that for the small amplitudes of real GWs can be approximated as 
\begin{equation}
    \psi_{00}^{(\delta L)} = \psi_{00}\left( 1+ i\phi(t) \right)\, ,
\end{equation}
where the extra phase is
\begin{equation}
    \phi(t) = \frac{4\pi}{\lambda}\delta L(t)\, .
\end{equation}

This GW generates some audio sidebands as 
\begin{equation}
    \vec{\mathbbm{s}}_{\rm e}^{~(\delta L)} = \begin{bmatrix}
        i\tilde{\phi}(\Omega) \\
        -i\tilde{\phi}(\Omega)
    \end{bmatrix}\, ,
\end{equation}
where we have used the fact that $\tilde{\phi}^*(-\Omega)=\tilde{\phi}(\Omega)$. This field perturbation in the two photon quadrature would be 
\begin{equation}
    \vec{\mathbbm{s}}_{\rm e}^{~(\delta L)}= \sqrt{2}\frac{4\pi}{\lambda}\delta L ~\ep \, ,
\end{equation}
which is purely on the phase quadrature. The corresponding readout for this would be
\begin{equation}
    y^{(\delta L)} = \sqrt{2}\frac{4\pi}{\lambda}\delta L \Tvec{e}^\dagger \ep\, ,
\end{equation}
which can be equated to Eq.~\eqref{eq:sediff} in order to find the equivalent displacement, yielding
\begin{equation}
    \delta L_{\rm eq}^{\rm (diff)} = \frac{\lambda}{4\pi}\frac{\Tvec{e}^\dagger \sdiff}{\Tvec{e}^\dagger\ep}\, .
\end{equation}

We are interested on the strain, $h=\delta L/L$\footnote{For long-arm detectors such as CE, displacement noise divided by the arm length is not, strictly speaking, equivalent to strain at frequencies comparable to, or exceeding, the free spectral range. In this regime, the displacement--strain conversion includes additional corrections that depend on frequency and on the source location~\cite{Rakhmanov:2008is,Essick:2017wyl}. Nonetheless, for the frequencies of interest to stray light noise, which are much smaller than the free spectral range, this is a good estimation of the strain.} and its PSD becomes  
\begin{equation}\label{eq:hdiff}
    \mathcal{S}_{h^{\rm(diff)}}(f)= \left(\frac{\lambda}{4\pi L_a}\right)^2\frac{\Tvec{e}^\dagger\mathbb{S}_{\rm diff}\Tvec{e}}{\left| \Tvec{e}^\dagger\ep\right|^2 }\, ,
\end{equation}
with 
\begin{equation}
    \mathbb{S}_{\rm diff}=
    \begin{bmatrix}
        \delta a_{R}^2 & \delta a_{R}\delta a_{I}\\
        \delta a_{R}\delta a_{I} & \delta a_{I}^2
    \end{bmatrix}\mathcal{S}_{x_\perp}(f)\, ,
\end{equation}
where $\mathcal{S}_{x_\perp}(f)$ denotes the PSD of the displacement $x_\perp(t)$.

Of special importance is to note that the matrix $\mathbb{S}_{\rm diff}$ contains non-vanishing off-diagonal terms, which can create some mixing between the amplitude and phase terms, the magnitude of which is controlled by the interferometer response encoded in $\Tvec{e}$. Finally, we can add the contributions from all $N_{\rm b}$ baffles as
\begin{equation}
    \mathcal{S}_{h^{(\rm diff)}}(f)=\sum_{j=1}^{N_{\rm b}}\mathcal{S}_{h^{(\rm diff)}_j}(f)\, ,
\end{equation}
where it is assumed that each one contributes independently to the noise. 

The equation employed in previous analysis (see for example Refs.~\cite{Andres-Carcasona:2023qom,Andres-Carcasona:2025xwq,Andres-Carcasona:2024hel}), based on the original works of Refs.~\cite{Thorne89,FlanaganThorne95_Diff,Flanagan94,Flanagan95,Vinet96,Vinet97,Brisson98} was 
\begin{equation}
     \mathcal{S}_{h^{\rm(diff)}}^{\rm legacy}(f)= \left(\frac{\lambda}{4\pi L_a}\right)^2\delta a_I^2~\mathcal{S}_{x_\perp}(f)\,.
\end{equation}

This is equivalent to taking only the phase component of the diffraction field as it is the one imprinting a change in the phase, equivalent to the effect of a GW passing by. It becomes evident that this ignores any radiation pressure caused by the real part, but also any cross-coupling. Furthermore, it does not take into account the effect by the extraction cavity or the readout scheme.

Using the approximate expressions of Sec.~\ref{sec:approximate} we can simplify Eq.~\eqref{eq:hdiff} under the typical conditions of signal extraction, no detunings and phase readout, to obtain
\begin{equation}\label{eq:Shdiff_approx}
    \mathcal{S}_{h^{\rm(diff)}}(f)\simeq \left( \frac{\lambda}{4\pi L_a} \right)^2\left(\delta a_I-\delta a_R\mathcal{K}\Gamma^{\rm RSE}\right)^2\mathcal{S}_{x_{\perp}}(f)\, .
\end{equation}
This expression shows the appropriate correction that must be applied in order to account for radiation pressure under the assumptions aforementioned. If the detector violates any of these assumptions then Eq.~\eqref{eq:Shdiff_approx} is no longer valid and one must fall back to Eq.~\eqref{eq:hdiff} with the full transfer factors.

\subsection{Backscattering}
The backscattering noise calculation usually starts by computing the fraction of power that recombines with the main beam, which we denote by
\begin{equation}
    \varepsilon \equiv \sqrt{\frac{\delta P}{P}}\, .
\end{equation}

This can be done analytically using the bidirectional reflectance distribution function (BRDF) of the mirror~\cite{Thorne89,FlanaganThorne95_Diff,Flanagan94,Flanagan95,Vinet96,Vinet97,Brisson98} or numerically using modern softwares~\cite{Andres-Carcasona:2023qom,Andres-Carcasona:2024hel,Andres-Carcasona:2025xwq,Romero-Rodriguez:2020bys,Romero-Rodriguez:2022mje}. This fraction of recoupled power can be estimated using Throne and Flanagan's formalism as~\cite{Flanagan94}
\begin{equation}
    \varepsilon^2 = \left(\frac{\lambda}{r} \right)^2 \left(\frac{\td \mathcal{P}}{\td \Omega_{\rm ms}}\right)^2\frac{\td \mathcal{P}}{\td \Omega_{\rm bs}}\delta \Omega_{\rm ms}\, ,
\end{equation}
where $r$ is the distance to the backscattering surface, $\td \mathcal{P}/\td \Omega_{\rm ms}$ the probability of a photon being scattered by the mirror in the solid angle $\td \Omega_{\rm ms}$ and $\td \mathcal{P}/\td \Omega_{\rm bs}$ the probability of a photon scattering at the backscattering surface in a solid angle $\td \Omega_{\rm bs}$. These probabilities are directly related to the BRDFs of both surfaces (mirror and backscatterer) for a normal incident light beam as 
\begin{equation}
    \mathrm{BRDF}(\theta)=\frac{1}{\cos\theta}\frac{\td \mathcal{P}}{\td \Omega_{\rm ms}}\, ,
\end{equation}
where $\theta$ denotes the azimuthal angle to the backscattering surface from the normal direction.

We consider the case where the scattered field recouples at the same mirror that has emitted the light. Current estimations, typically compute the scattered light from the ITM that comes back and double the result as the cavity is symmetrical.  Therefore, the source vector is defined at the node $\avec{iho}$, and the appropriate transfer factor is $\Tvec{i}^\dagger$, as derived in Sec.~\ref{sec:TFs}.

The recoupled backscattered field can be written as a complex fractional field
\begin{equation}
    \delta a_{\rm bs}(t)=\varepsilon\,e^{i\Phi(t)},
\end{equation}
where \(\Phi(t)\) is the optical phase accumulated by the scattered path. In frequency domain this extra field becomes, in the sideband picture,
\begin{equation}
    \sbs(\Omega) = \varepsilon\begin{bmatrix}
        \tilde{C}(+\Omega)+i\tilde{S}(+\Omega) \\
        \tilde{C}^*(-\Omega)-i\tilde{S}^*(-\Omega)
    \end{bmatrix}\, ,
\end{equation}
where $\tilde{C}$ and $\tilde{S}$ denote the Fourier transforms of $\cos(\Phi(t))$ and $\sin(\Phi(t))$, respectively. This can be translated into the amplitude and phase quadratures resulting in 
\begin{equation}
    \sbs = \sqrt{2}\varepsilon\begin{bmatrix}
        \tilde{C}(\Omega) \\
        \tilde{S}(\Omega) 
    \end{bmatrix}\, ,
\end{equation}
where since the sine and cosine functions are real, we have used the fact that $ \tilde{C}^*(-\Omega)= \tilde{C}(\Omega)$ and $ \tilde{S}^*(-\Omega)= \tilde{S}(\Omega)$. This expression shows that, in general, backscattering injects both amplitude and phase quadratures. 
The readout produced by backscattering is then
\begin{equation}
   y^{\rm (bs)}=
    \Tvec{i}^\dagger\,\sbs\, .
\end{equation}

To convert this perturbation into an equivalent displacement, we compare it to the readout produced by an arm length change $\delta L$
\begin{equation}
    y^{(\delta L)}=\sqrt{2}\frac{4\pi}{\lambda}\delta L\Tvec{i}^\dagger\ep\, ,
\end{equation}
the equivalent displacement is
\begin{equation}
    \delta L_{\rm eq}^{\rm (bs)}=
    \frac{\lambda}{4\pi}
    \frac{\Tvec{i}^\dagger\sbs}
         {\Tvec{i}^\dagger\ep}\, .
    \label{eq:dLeq_bs_general_rewrite}
\end{equation}

We are primarily interested in the phase acquired due to motion of the backscattering surface along the photon propagation direction~\cite{Flanagan94}. Denoting by $x_{\parallel}(t)$ the parallel motion with respect to the optical path, the additional phase is
\begin{equation}
    \Phi(t)=\frac{4\pi}{\lambda}x_{\parallel}(t)\, .
\end{equation}

The dependence on this phase is nonlinear in $x_{\parallel}(t)$ because of the sine and cosine terms and, therefore, phase-wrapping (or upconversion) can be important. 

From Eq.~\eqref{eq:dLeq_bs_general_rewrite}, we can compute the PSD of the backscattering strain noise, which yields
\begin{equation}
    \mathcal{S}_{h^{(\rm bs)}}(f)=
    \left(\frac{\lambda\varepsilon}{4 \pi L_a}\right)^2
    \frac{\Tvec{i}^\dagger\mathbb{S}_{\rm bs}\,\Tvec{i}}
         {\left|\Tvec{i}^\dagger\ep\right|^2}\, .
    \label{eq:hbs_matrix_final}
\end{equation}
where the cross-power spectral density matrix is 
\begin{equation}
    \mathbb{S}_{\rm bs}(f)=
    \begin{pmatrix}
        \mathcal{S}_{cc}(f) & \mathcal{S}_{cs}(f)\\
        \mathcal{S}_{sc}(f) & \mathcal{S}_{ss}(f)
    \end{pmatrix}.
\end{equation}
and each term defined as 
\begin{equation}
    \mathcal{S}_{cc}(f)=\mathrm{PSD}[C(t)],\,\,\,
    \mathcal{S}_{ss}(f)=\mathrm{PSD}[S(t)],
\end{equation}
\begin{equation}
    \mathcal{S}_{cs}(f)=\mathrm{CPSD}[C(t),S(t)],\,\,\,
    \mathcal{S}_{sc}(f)=\mathrm{CPSD}[S(t),C(t)]\, ,
\end{equation}
where CPSD denotes the cross-power spectral density of the two time series.  The off-diagonal terms, account for correlations between the cosine and sine quadratures induced by the common phase process $\Phi(t)$, and are in general non-zero. The nonlinear phase-wrapping of backscattered light noise has been experimentally observed in LIGO~\cite{Ottaway:2012oce} and Virgo~\cite{Was:2020ziy} and, for this reason, we keep the full phase-wrapped expression.

Finally, the contributions from all \(N_{\rm b}\) baffles can be added in quadrature as
\begin{equation}
    \mathcal{S}_{h^{(\rm bs)}}(f)=\sum_{j=1}^{N_{\rm b}}\mathcal{S}_{h^{(\rm bs)}_j}(f)\, ,
\end{equation}
where the sum assumes that the contributions from different baffles are statistically independent.

The model employed in previous analysis (see for example Refs.~\cite{Thorne89,FlanaganThorne95_Diff,Flanagan94,Flanagan95,Vinet96,Vinet97,Brisson98,Andres-Carcasona:2023qom,Andres-Carcasona:2025xwq,Andres-Carcasona:2024hel,Hiro_ETMripple,Romero-Rodriguez:2022mje,Macquet:2022simsVirgo}) was 
\begin{equation}\label{eq:hbs_legacy}
     \mathcal{S}_{h^{\rm(bs)}}^{\rm legacy}(f)= \left(\frac{\lambda\varepsilon}{4\pi L_a}\right)^2\left[ 1+\left( \frac{8\Gamma^{\rm RSE} P}{c\lambda m\pi f^2}\right)^2\right]~\mathcal{S}_{ss}(f)\,.
\end{equation}

Using the approximate expressions of Sec.~\ref{sec:approximate} we can simplify Eq.~\eqref{eq:hbs_matrix_final} under the typical conditions of signal extraction, no detunings and phase readout, to obtain
\begin{multline}\label{eq:Shbs_approx}
    \mathcal{S}_{h^{\rm(bs)}}(f)\simeq \left( \frac{\lambda\varepsilon}{4\pi L_a} \right)^2\left [\left(\mathcal{K}\Gamma^{\rm RSE}\right)^2\mathcal{S}_{cc}+\mathcal{S}_{ss}\right.\\ \left. -2\mathcal{K}\Gamma^{\rm RSE}\mathrm{Re}\left(\mathcal{S}_{cs}\right) \right]\, .
\end{multline}
In practice, for a realistic motion spectra, $\mathcal{S}_{cc}\simeq \mathcal{S}_{ss}>|\mathcal{S}_{cs}|$, which reduces even further to Eq.~\eqref{eq:hbs_legacy}. Nonetheless, if the detector violates any of these aforementioned assumptions then Eq.~\eqref{eq:Shbs_approx} is no longer valid and one must fall back to Eq.~\eqref{eq:hbs_matrix_final} with the full transfer factors.

\section{Numerical results}\label{sec:Results}
In this section, we present some numerical results using the extended equations for LIGO, CE and ET-LF representative parameters as reported in Tab.~\ref{tab:variables_numbers} and a free mass susceptibility model\footnote{These values are intended to be representative interferometer configurations for transfer-factor comparisons, not as finalized detector design specifications.}. 

\begin{table}[htbp]
    \centering
    \begin{tabular}{c|c|c|c|c}
       \hline
       \textbf{Variable} & \textbf{Value LIGO} & \textbf{Value CE} & \textbf{Value ET-LF} & \textbf{Units} \\
       \hline\hline
       $\mathcal{T}_e$ & 5 & 5 & 5 & [ppm] \\
       \hline
       $\mathcal{L}_e$ & 35 & 20 & 20 & [ppm] \\
       \hline
       $\mathcal{T}_i$ & 1.40 & 1.40 & 1.38 & [\%] \\
       \hline
       $\mathcal{L}_i$ & 35 & 16 & 13 & [ppm] \\
       \hline
       $\mathcal{T}_s$ & 32.5 & 2.0 & 20.0 & [\%] \\
       \hline
       $\mathcal{L}_s$ & 200 & 7 & 13 & [ppm] \\
       \hline
       $\lambda$ & 1064 &  1064 & 1550 & [nm] \\ \hline
       $P$ & 0.4 & 1.5 & 0.018 & [MW] \\ \hline
       $L_a$ & 4 & 40 & 10 & [km] \\ \hline
       $L_s$ & 55 & 117 & 100 & [m] \\ \hline 
       $m$ & 40 & 440 & 211 & [kg] \\ \hline
       $\varphi_a$ & 0 & 0 & 0 &  [rad] \\ \hline
       $\varphi_s$ & $\pi/2$ & $\pi/2$ & $\pi/2+0.6$ & [rad] \\ \hline
       $\zeta$ & $\pi/2$ & $\pi/2$ & $\pi/2+0.6$ & [rad] \\ \hline \hline
    \end{tabular}
    \caption{Baseline parameters assumed for the numerical tests unless stated otherwise. The terms $\mathcal{T}=t^2$ indicates the power transmittance and $\mathcal{L}=1-t^2-r^2$ the losses of the optics. Parameters from Refs.~\cite{LIGOScientific:2014pky,Evans:2021gyd,Hild11,ETdesign,pygwinc}}
    \label{tab:variables_numbers}
\end{table}

\subsection{Baseline transfer factors}\label{sec:baseline_TFs}
\begin{figure*}[htbp]
    \centering
    \includegraphics[width=1.0\linewidth]{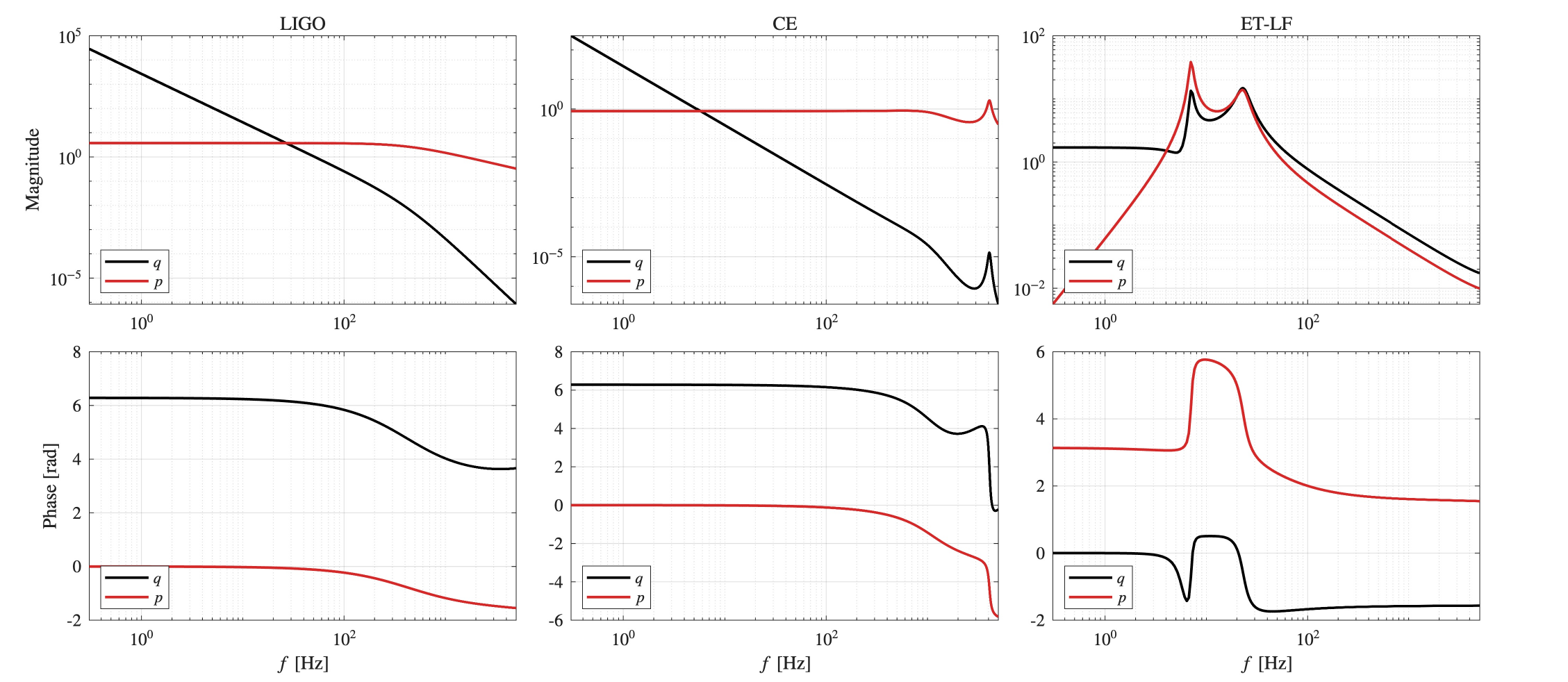}
    \caption{Amplitude and phase of the baseline transfer factors $\Tvec{e}$ for both quadratures for LIGO, CE and ET-LF.}
    \label{fig:BaselineTFs}
\end{figure*}
The two main components required to estimate both the backscattering and diffraction noises are $\Tvec{i}$ and $\Tvec{e}$. The baseline parameters described in Tab.~\ref{tab:variables_numbers} represent the detectors in signal extraction mode, where the readout homodyne angle is set to the same signal extraction cavity detuning, $\varphi_s = \zeta = \pi/2$. Nonetheless, ET-LF is expected to operate with a detuned signal extraction cavity and this will substantially affect the transfer factor of stray light. Although $\Tvec{i}$ includes an additional ETM reflection and an extra arm propagation ($\Lmat{a}$) compared to $\Tvec{e}$, the arm cavity's high finesse means a scattered photon bounces hundreds of times before exiting. This massive optical gain makes the macroscopic coupled-cavity dynamics dominate, rendering the single-trip differences essentially negligible. Therefore, $\Tvec{e}$ and $\Tvec{i}$, are qualitatively and quantitatively very similar. The magnitude and phase of $\Tvec{e}$ obtained for these detectors are displayed in Fig.~\ref{fig:BaselineTFs}.

Because of the chosen parameters, the readout is primarily sensitive to the phase quadrature of the field, as it is usual for gravitational wave detection. This explains why the magnitude of the $p$-component is approximately unity across the sensitive band. A pure phase perturbation injected into the arm couples almost directly to the final phase-sensitive measurement. For ET-LF, this behavior is slightly different, but the main reason lies again in the detuning. The target of ET-LF is to be very sensitive in the band $\sim3-30$~Hz instead of broadband, and this is seen in the transfer factors as the region for which the phase readout is enhanced.

Conversely, an amplitude perturbation would theoretically not be measured by a pure phase readout in a cavity. However, its low-frequency contribution is significantly enhanced, decreasing rapidly with frequency according to an $f^{-2}$ scaling for the tuned case such as in LIGO or CE. This is the classic scaling of radiation pressure quadrature. For ET-LF, this scaling is broken and again the reason is in the signal extraction cavity detuning. Unlike for tuned cases, radiation pressure can couple better than phase perturbations at both low and high-frequencies, but outside of the target detection band. Interestingly enough, at low frequencies remains roughly constant and at higher frequencies decays with a scaling of $f^{-1}$.

At higher frequencies, where the radiation-pressure effect becomes negligible for LIGO and CE, the phase response dominates. In the kHz region, the dynamics of the cavity introduce a dip followed by a narrow resonance-like feature in CE's transfer factors. This is a direct consequence of the long arm cavities as the free spectral range (FSR) becomes $c/(2L_a) \approx 3.75$~kHz, which is within the detection band.

Overall, Fig.~\ref{fig:BaselineTFs} demonstrates that while the approximations used until now showed the overall appropriate scalings, these were only good approximations for the case with signal extraction and no detunings. Additionally, they show how diffraction can also contribute to the noise coupling via radiation pressure, something that some earlier studies did not account for~\cite{Thorne89,VajenteCEbaffles}.  Finally, they show how the signal extraction cavity design does have an important effect on the estimation of stray light, as it noticeably changes how perturbation in the main arms couple to the final readout.

\subsection{Comparison of diffraction noise}
An important point is to quantify the effect on the diffraction noise estimation with the derived method when compared to the legacy one. To do so we can evaluate the ratio between the two curves  
\begin{equation}
    \mathcal{R}_{\rm diff}(f) \equiv \frac{\mathcal{S}_{h^{\rm(diff)}}(f)}{\mathcal{S}_{h^{\rm(diff)}}^{\rm legacy}(f)}\, .
\end{equation}

In order to cover a representative parameter space we can assume that the perturbation is of the form 
\begin{equation}
    \sdiff = |\delta a|e^{i\theta}\sqrt{\mathcal{S}_{x_\perp}(f)}\, ,
\end{equation}
such that the phase and amplitude content are governed by the angle $\theta$. The legacy method, on the contrary, assumes that all the perturbation lives in the phase quadrature. Taking the quotient removes the dependence on the magnitude of the perturbation $|\delta a|$ and the vibration spectra, such that $\mathcal{R}_{\rm diff}$ is mainly a function of frequency and the angle $\theta$.

\begin{figure*}[htbp]
    \centering
    \includegraphics[width=1.0\linewidth]{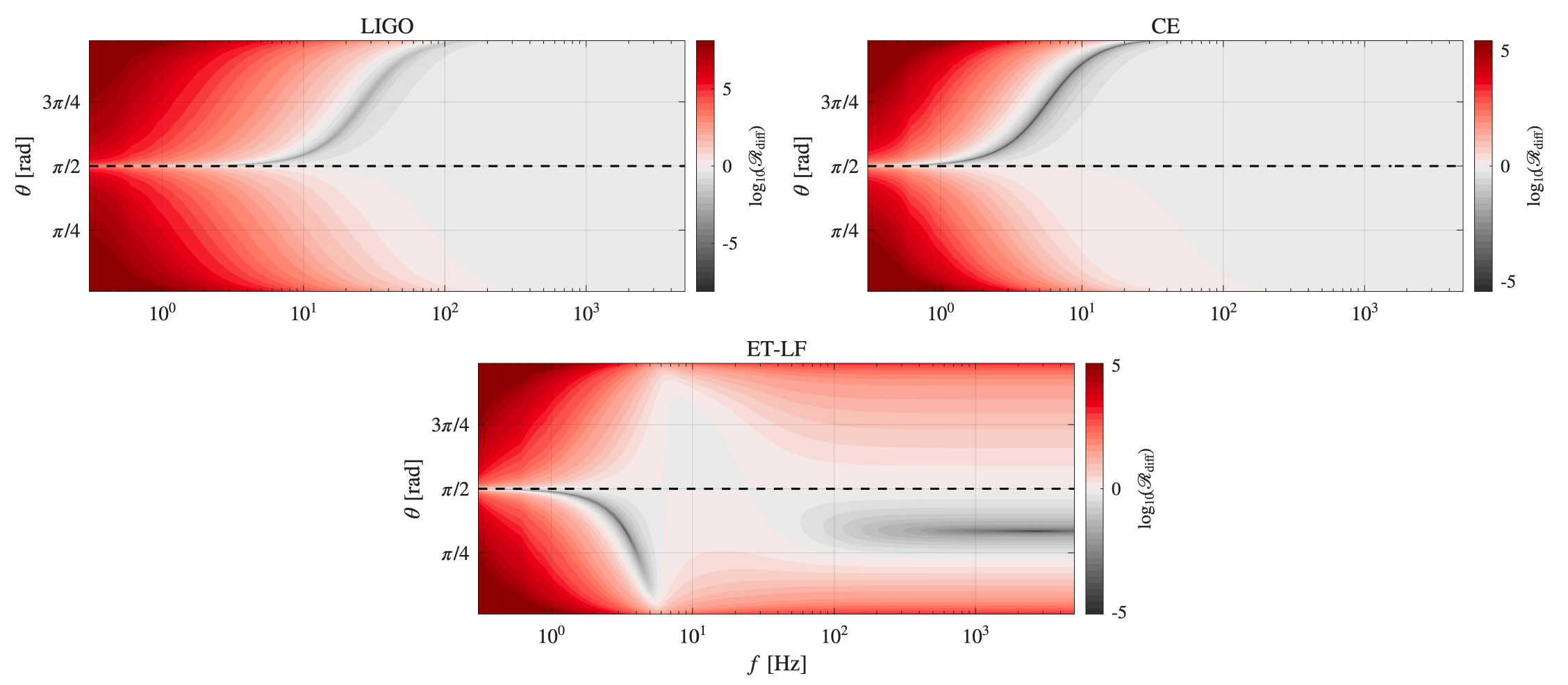}
    \caption{Contour plot of the quotient of the model derived in this work and the previous literature one, $\mathcal{R}_{\rm diff}$, as a function of the frequency and the diffraction perturbation angle $\theta$. The black dashed line shows the contour where both models are the same ($\mathcal{R}_{\rm diff}=1$).}
    \label{fig:BaselineRdiff}
\end{figure*}

In Fig.~\ref{fig:BaselineRdiff} we show the values of $\mathcal{R}_{\rm diff}$ for the three detectors considered. For a purely imaginary perturbation, corresponding to $\theta=\pi/2$, both models agree perfectly ($\mathcal{R}_{\rm diff}=1$), which is a good sanity check. Nonetheless, when the diffraction perturbation starts having a real part (and hence $\theta\neq \pi/2$), both models start disagreeing by several orders of magnitude. The reason is that current models ignore or do not appropriately treat the radiation pressure part, which can translate into a phase-noise as shown by the transfer factors. In the case of ET-LF, there is the additional  contribution of the detuning, which modifies the frequency response of both phase and amplitude noises. 

At frequencies below $\sim60$~Hz for LIGO ($\sim 20$~Hz for CE), the plot is dominated by large values of $\mathcal{R}_{\rm diff}$. As demonstrated in Sec.~\ref{sec:baseline_TFs}, the amplitude transfer factor scales as $f^{-2}$ due to the radiation-pressure back-action on the test masses. Unless the coupling is perfectly imaginary ($\theta = \pi/2$), this highly amplified amplitude-to-phase conversion overwhelms the direct phase contribution. Consequently, the legacy model severely underestimates the low-frequency noise for almost any generic diffraction coupling. This same effect is present for ET-LF, although it appears both above and below its most sensitive bandwidth of $\sim 3-30$~Hz. This is a direct consequence of the amplitude quadrature transfer factor component being above the phase one, as shown in Fig.~\ref{fig:BaselineTFs}, which enhances the radiation pressure coupling.  

Another interesting feature in Fig.~\ref{fig:BaselineRdiff} are the deep suppression bands that trace through the parameter space. Inside these regions, $\mathcal{R}_{\rm diff} < 1$, meaning that the full model predicts a lower noise than the legacy one. This occurs due to destructive interference. Depending on the complex angle $\theta$, the direct phase perturbation and the radiation-pressure-induced phase shift can have opposite signs. Along the darkest contour, these two contributions perfectly cancel each other out, creating an optomechanical blind spot to diffraction noise for the baseline specific geometry of each of the detectors.

For LIGO and CE, above $\sim 100$~Hz, the mechanical susceptibility of the test masses drops significantly, suppressing the radiation-pressure cross-coupling. Consequently, the ratio $\mathcal{R}_{\rm diff}$ approaches unity for most values of $\theta$, meaning the legacy assumption is generally valid at higher frequencies. This is not valid for the ET-LF because of the detuned signal extraction cavity.

\subsection{Comparison of backscattering noise}

\begin{figure*}[htbp]
    \centering
    \includegraphics[width=1.0\linewidth]{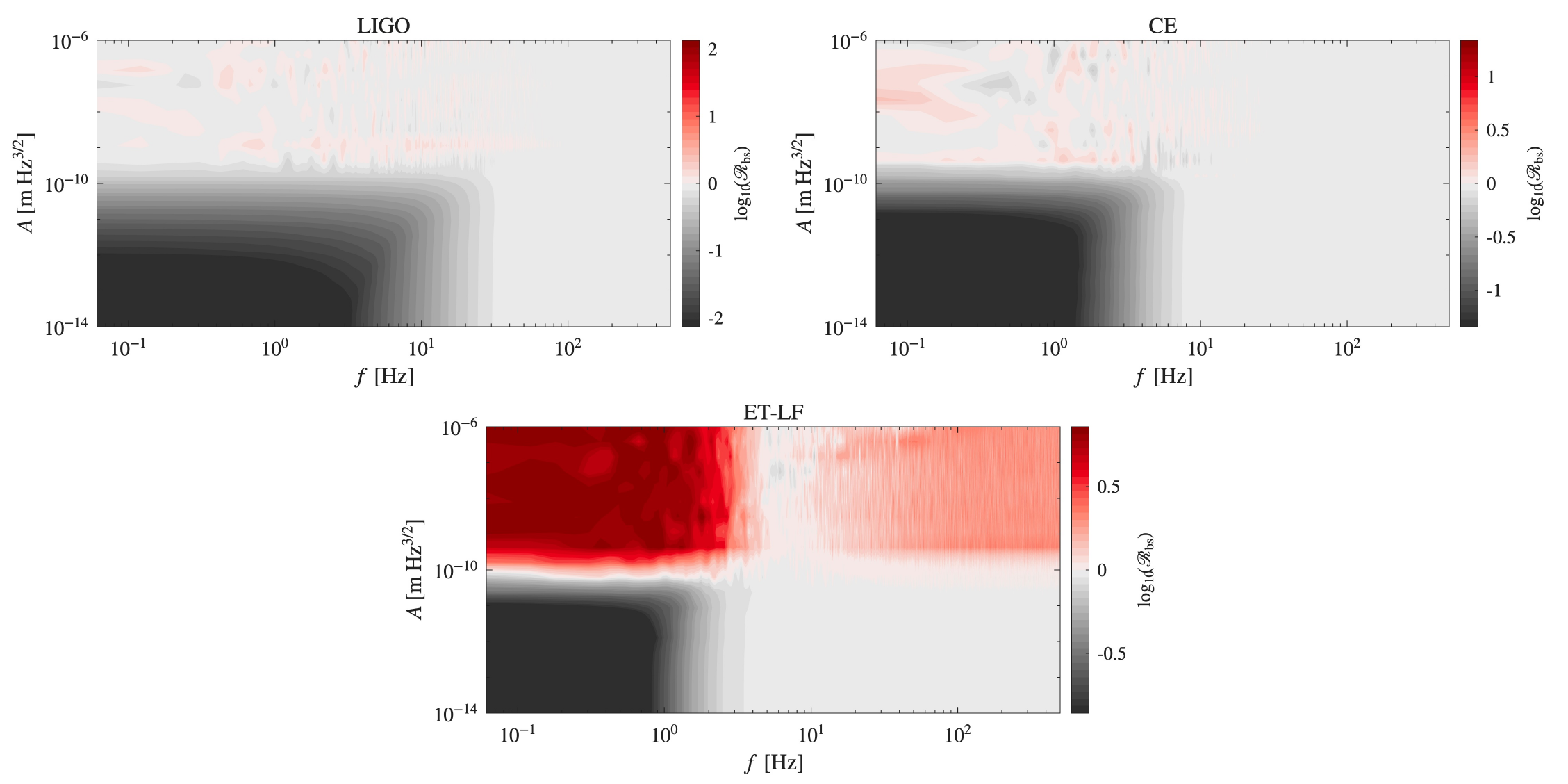}
    \caption{Contour plot of the quotient of the model derived in this work and the previous literature one, $\mathcal{R}_{\rm bs}$, as a function of the frequency and the amplitude of the spectral motion, $A$, when the motion ASD is assumed to be of the form $A/f^2$.}
    \label{fig:BaselineRbs}
\end{figure*}

We perform a similar comparative analysis for the backscattering noise by defining the ratio between the full transfer factor estimate and the legacy model as
\begin{equation}
    \mathcal{R}_{\rm bs}(f, A) \equiv \frac{\mathcal{S}_{h^{\rm(bs)}}(f)}{\mathcal{S}_{h^{\rm(bs)}}^{\rm legacy}(f)}\, .
\end{equation}
Because backscattering noise heavily depends on the non-linear phase-wrapping of the optical path, this ratio is evaluated as a function of both frequency $f$ and the amplitude of the backscatterer's motion, $A$, when the ASD of the motion is assumed to be of the form $A/f^2$. In Fig.~\ref{fig:BaselineRbs}, we present the contour map of $\log_{10}(\mathcal{R}_{\rm bs})$ for the three considered interferometers.


The most prominent feature in Fig.~\ref{fig:BaselineRbs} is the deep suppression region in the bottom-left corner, corresponding to low frequencies ($f \lesssim 10$~Hz) and small vibration amplitudes ($A \lesssim 10^{-10}$~m/Hz$^{3/2}$). In this regime, the optical phase accumulation is very small ($\Phi(t) \ll 1$), which implies that the field is almost entirely injected into the phase quadrature ($\sin\Phi \approx \Phi$) and the amplitude quadrature contains only a static DC component ($\cos\Phi \approx 1$) with negligible AC fluctuations at the fundamental frequency $f$. 

Consequently, our full model correctly predicts that, in this regime, backscattering does not induce AC radiation-pressure noise, leaving the overall noise unamplified at low frequencies. In contrast, the legacy model, indiscriminately applies a constant optomechanical gain factor to the entire noise spectrum. By artificially forcing this radiation pressure that does not physically exist for small linear vibrations, the legacy model overestimates the noise, causing the exact model ratio $\mathcal{R}_{\rm bs}$ to drop by orders of magnitude.

As the vibration amplitude $A$ increases, the assumption $\Phi(t) \ll 1$ breaks down. Phase wrapping spreads the vibrational energy across multiple harmonics and, crucially, this process populates the cosine quadrature with a broad AC spectrum. Once AC noise enters the amplitude quadrature, it couples to the test masses via radiation pressure, and the $f^{-2}$ optomechanical amplification naturally appears in our model. This is visually evident in Fig.~\ref{fig:BaselineRbs} as the dark dip abruptly vanishes at higher amplitudes, and the ratio approaches values closer to unity for LIGO and for CE.

In the upper half of the parameter space (large amplitude), the contour plot exhibits a highly structured, noisy speckle pattern. This is not a numerical artifact, but rather a physical consequence of quadrature mixing. When the backscattered phase is heavily wrapped, the sine and cosine quadratures become correlated and our matrix formalism explicitly incorporates them through the cross-power spectral densities ($\mathcal{S}_{cs}$ and $\mathcal{S}_{sc}$). This leads to some spectral interference that the legacy model ignores by only using the sine spectral density. In any case, the resulting fluctuations in $\mathcal{R}_{\rm bs}$ are of order unity for CE and LIGO, implying that the methods agree. Nonetheless, for  the ET-LF detector, again due to the detuning, the new model predicts almost an order of magnitude of more backscattering noise than the legacy, being most of the difference appreciable in the radiation pressure quadrant (high amplitude vibration and low frequency).

\subsection{Dependence on interferometer configuration}

The legacy analytical models for scattered-light noise couplings approximate the interferometer response as a static scalar gain. These expressions implicitly assume a fixed, perfectly resonant optical geometry and a pure phase-sensitive readout. However, advanced observatories are dynamically complex instruments. They routinely operate with detuned homodyne readout angles to optimize squeezed-light injection, actively tune their signal-recycling cavities to shape the bandwidth, and apply microscopic arm detunings (DARM offsets) to enable DC readout schemes. To accurately predict scattered light under these tunable states, the noise model must embed these macroscopic configuration parameters.

\begin{figure*}[htbp]
    \centering
    \includegraphics[width=\linewidth]{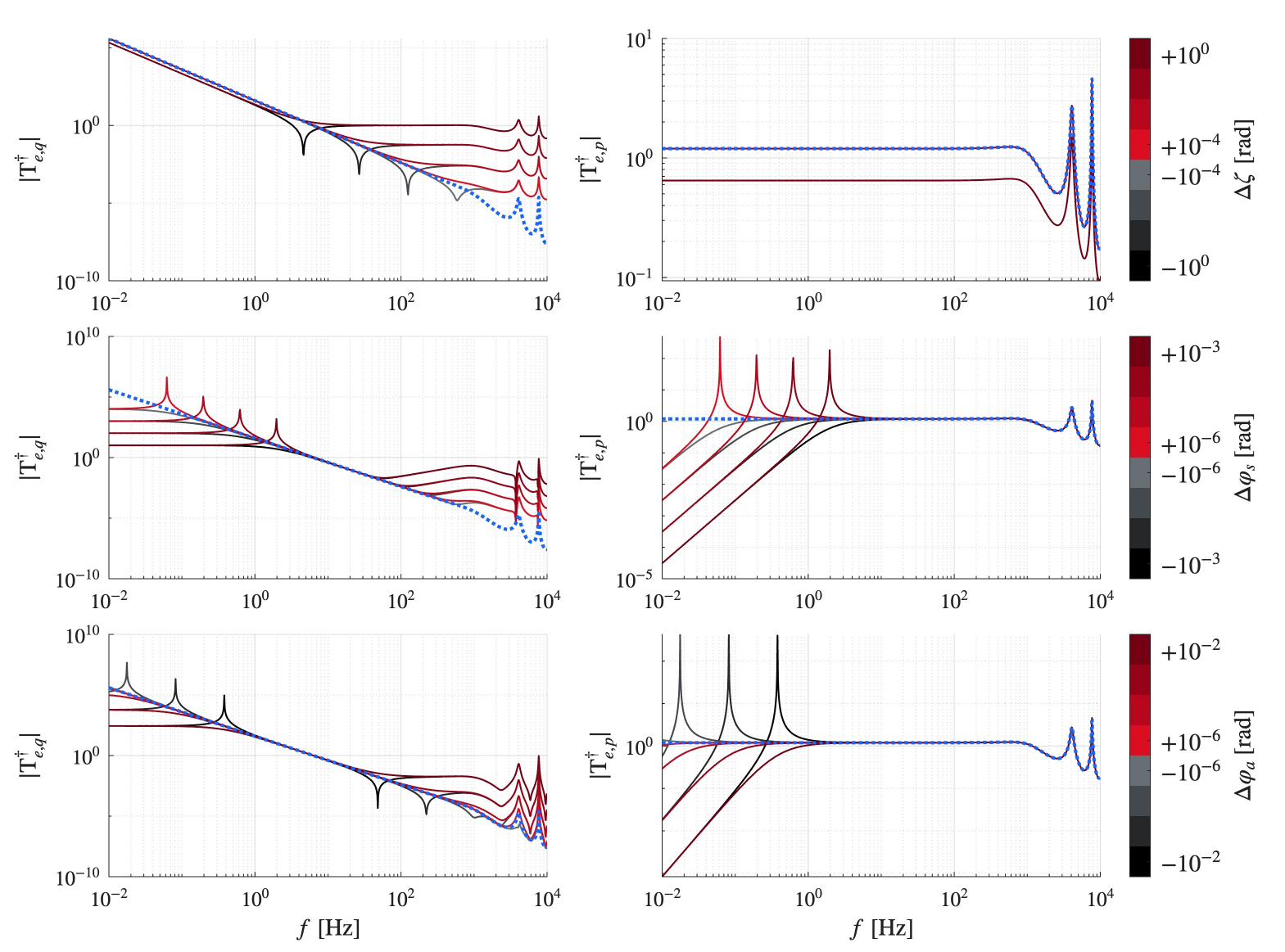}
    \caption{Magnitude of the transfer factor $\Tvec{e}$ for the amplitude (left) and phase (right) quadratures for CE when a change with respect to the baseline of the homodyne angle (top row), signal extraction cavity angle (middle row) or main arm detuning (bottom row) is present. The blue dashed line represents the baseline configuration.}
    \label{fig:ConfigSweep}
\end{figure*}

Using the equations derived, we systematically explore in this section the dependence of the scattered-light transfer factors on the interferometer configuration for CE (we focus only on this detector as a representative example). In Fig.~\ref{fig:ConfigSweep}, we present a sweep of the magnitude of the amplitude and phase components of the ETM transfer factor ($|\Tvec{e}|$). The nominal CE baseline transfer factor (as shown in Fig.~\ref{fig:BaselineTFs}) is represented by the dotted blue line. In each row, a single parameter is symmetrically varied around this baseline value.

In the top row of Fig.~\ref{fig:ConfigSweep}, we introduce a change $\Delta\zeta$ to the homodyne readout angle with respect to the baseline value shown in Tab.~\ref{tab:variables_numbers}. Because of this offset, any deviation from a pure phase measurement ($\zeta \neq \pi/2$) allows the amplitude quadrature ($\Tvec{e,q}^\dagger$) to couple directly into the GW signal channel. For positive offsets, the high-frequency end of the transfer factor increases and stays roughly constant after the optical spring frequency. This frequency gets smaller as the offset increases. On the other hand, if the offset is negative, there is optical anti-spring that creates a dip in the response at the resonant frequency. In both cases the low frequency $f^{-2}$ scaling is maintained. The phase response, on the other hand, is mostly unaffected. It requires a large and unrealistic offset of $1$~rad to produce a noticeable change of about a factor $2$.

These results show that the readout angle could theoretically be tuned to actively create a blind spot to suppress specific low-frequency scattered-light peaks arising from radiation pressure, a mitigation strategy entirely inaccessible to standard legacy noise models. In any case, this angle will be chosen according to other optical design considerations and not the other way around, but as shown, these choices will have an impact on scattered light noise estimations.

In the middle row of Fig.~\ref{fig:ConfigSweep}, we sweep the microscopic detuning of the signal extraction mirror by generating a small offset, $\Delta\varphi_s$, from the reference value reported in Tab.~\ref{tab:variables_numbers}. The signal cavity tuning dictates the resonant dynamics of the coupled detector and slightly detuning the signal cavity from pure resonance introduces an optical spring effect as sharp, undamped peaks. When the offset is negative the sharp resonance peak disappears and the spring is overdamped. In both cases, the radiation pressure component is lowered at small frequencies and enhanced at large frequencies and the phase component is diminished below the optical spring frequency. Because the legacy models treat the arm and signal cavity interactions as a flat, frequency-independent scalar gain, they can miss some of these optomechanical dynamical shifts.

Finally, in the bottom row of Fig.~\ref{fig:ConfigSweep}, we introduce a microscopic detuning to the main arm cavities, $\Delta\varphi_a$, characteristic of the DARM offset required for DC readout. Similar to the signal cavity tuning, a DARM offset introduces a powerful optical spring, generating moving resonant peaks at low frequencies. Similarly, it reduces the low frequency radiation noise and phase noise components, while it increases the high-frequency radiation pressure one. For positive detuning, the oscillations are damped, while for negative detunings the resonant peaks appear. Interestingly, there are also anti springs in the radiation pressure component that appear in the detection band.

Altogether, Fig.~\ref{fig:ConfigSweep} demonstrates that the scattered-light coupling is not a static quantity as previously assumed, but rather a quantity that shifts continuously as the instrument is being tuned. Neglecting these properties leads to an incomplete picture of these couplings.

\subsection{Revisiting ET-LF noise estimation}

A final result that can be looked at is the difference that the modeling here presented produces in the estimation of the ET-LF stray light noise budget. The three main changes with respect to Ref.~\cite{Andres-Carcasona:2025xwq} are: the no-upconversion of the noise curve for diffraction, the improved transfer factors and the correct treatment of the radiation pressure of diffraction. The simulation parameters for SIS and baffle configuration are the same as the ones presented in Ref.~\cite{Andres-Carcasona:2025xwq} to allow a fair comparison. 

\begin{figure}[htbp]
    \centering
    \includegraphics[width=1.0\linewidth]{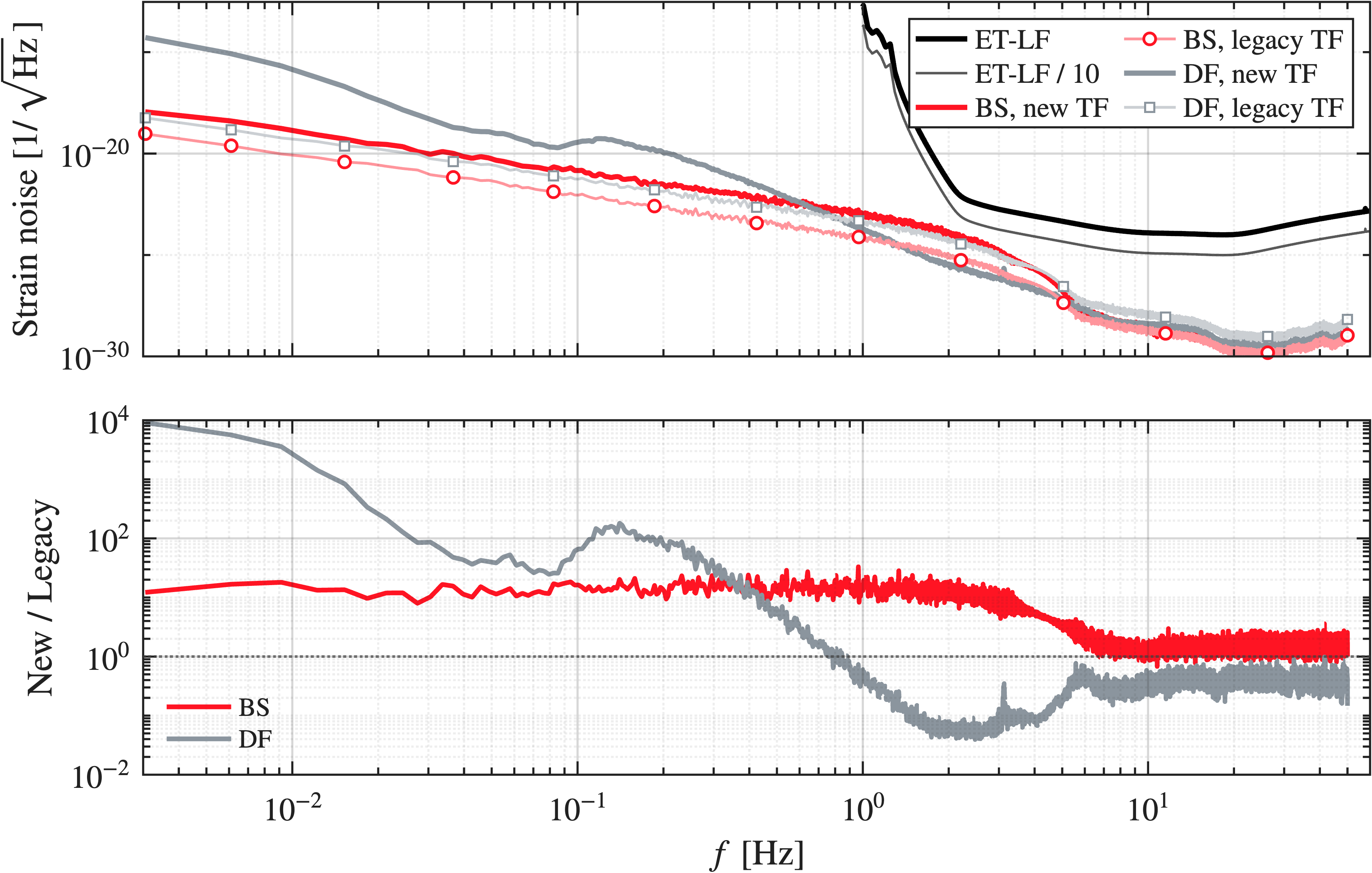}
    \caption{ (\textit{upper}) Backscattering (BS) and diffraction (DF) noises for ET-LF detector with the simulation parameters of Ref.~\cite{Andres-Carcasona:2025xwq} with the legacy and new models alongside the ET-LF noise curve and the safety margin of an order magnitude below (\textit{lower}) ratio of the new and legacy noise estimations.}
    \label{fig:NoiseCompETLF}
\end{figure}

We present in Fig.~\ref{fig:NoiseCompETLF} the comparison between the legacy model and the one derived in this work. The backscattering noise is about an order of magnitude larger with the new model up to about $4$~Hz, which is the SQL frequency, as shown also in the TFs of Fig.~\ref{fig:BaselineTFs}. From that frequency on, the difference is mostly negligible, indicating that the models mostly agree, at least up to $50$~Hz, the limit for which the seismic data is available.

\begin{figure}[htbp]
    \centering
    \includegraphics[width=1.0\linewidth]{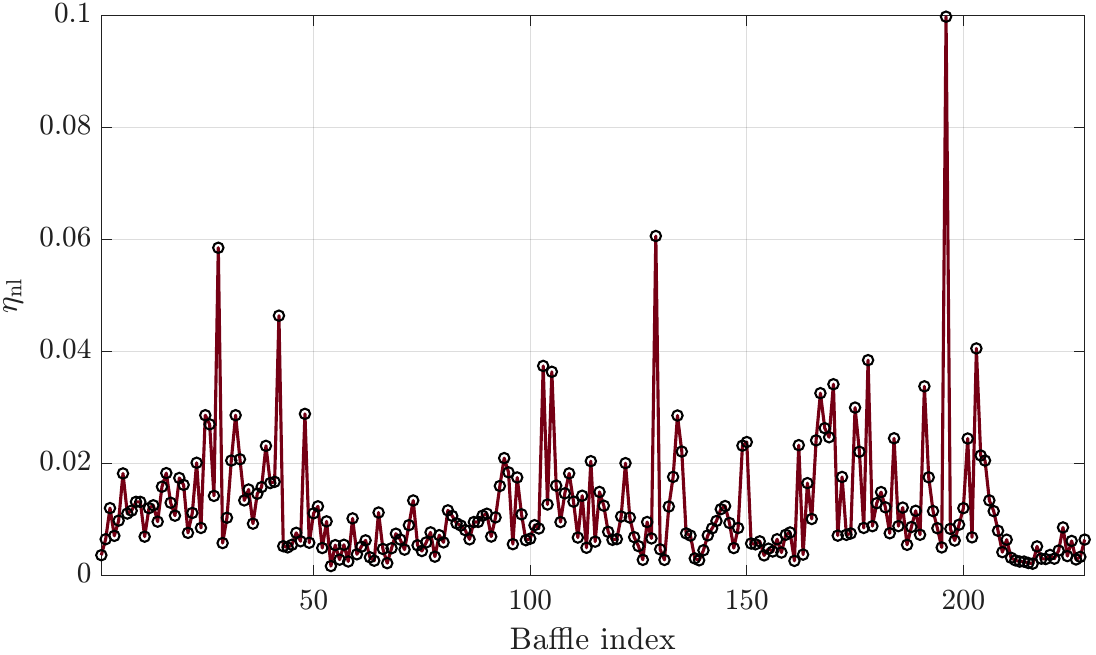}
    \caption{Relative non-linear contribution as defined in Eq.~\eqref{eq:eta_nl} for the ET-LF noise estimation.}
    \label{fig:eta_nl}
\end{figure}

The diffraction noise, on the other hand, is much larger than originally predicted up to a frequency of about $1$~Hz. This is the contribution of radiation pressure of diffraction noise, which the legacy models ignore\footnote{We note that Refs.~\cite{Andres-Carcasona:2023qom,Andres-Carcasona:2025xwq} included a correction to account for the radiation pressure in the diffraction noise. In particular, they apply the same prescription as for backscattering noise with the static gain of the SEM-ITM cavity. Nonetheless, it was applied to the phase quadrature perturbation instead of the amplitude one, as it targeted an order of magnitude estimation rather than a precise derivation of the effect of radiation pressure in diffraction noise.}. Above $1$~Hz there is a noticeable dip that is mostly related to the non-phase wrapped vibration spectrum that we use in the new model instead of the phase-wrapped one of Refs.~\cite{Andres-Carcasona:2023qom,Andres-Carcasona:2025xwq}. In such references it was assumed that the same non-nonlinearities that affected backscattering could affect diffraction, but using the formalism of Appendix~\ref{app:phase_wrapping_diffraction} we can see, as shown in Fig.~\ref{fig:eta_nl}, that the nonlinear term is subdominant across all baffles. In fact, only one baffle reaches a $\eta_{\rm nl} \sim 0.1$, being all the others below this value. This shows that for this case, the linearity assumption of diffraction coupling holds across all baffles and no phase-wrapping or equivalent is necessary.

Overall, these results show that the accurate modeling of radiation pressure, accounting for the correct transfer factors and the nonlinearities of the vibration spectrum are necessary as ignoring them can lead to non-negligible changes in the stray light noise estimations. The transfer factor part is especially important for ET-LF which operates with the detuned signal extraction cavity.

\section{Conclusions}

In this work, we have improved some modeling aspects of stray-light noise estimation inside of the main arms of ground-based GW detectors. For that, we have used a set of optomechanical transfer factors to convert scattered light perturbations in the main arms into equivalent strain noise. This approach makes it possible to propagate generic amplitude and phase perturbations from the FP arm to the final readout while consistently accounting for radiation-pressure effects, signal extraction dynamics, microscopic detunings, and the homodyne readout angle.

These transfer factors show that scattered light noise cannot, in general, be described by a static scalar coupling as has been done to guide the baffles in the beamtubes of next-generation detectors~\cite{Andres-Carcasona:2023qom,Andres-Carcasona:2025xwq}. Instead, its contribution to the measured strain depends on the full optomechanical response of the detector. This is particularly relevant whenever the scattered field contains both amplitude and phase quadrature components, as occurs generically for diffraction and backscattering. In tuned configurations representative of LIGO and CE, the legacy expressions reproduce the correct qualitative behavior in regimes dominated by phase perturbations, especially at high frequencies where radiation-pressure effects are suppressed. However, at low frequencies the amplitude quadrature can be strongly enhanced through radiation pressure, producing sizeable deviations from previous estimates whenever the scattered perturbation is not purely phase-like.

For diffraction noise, this formalism shows that the real part of the diffraction-induced field perturbation can contribute to the strain through radiation-pressure coupling. This effect is absent, or only approximately treated, in previous models that retained only the imaginary part of the perturbed overlap coefficient. As a result, the legacy model can significantly underestimate diffraction noise at low frequencies for generic complex diffraction couplings. At the same time, the full transfer factor treatment also reveals regions of destructive interference where the direct phase contribution and the radiation pressure contribution partially cancel, leading to reduced coupling relative to the legacy estimate. These optomechanical blind spots are naturally captured by the formalism here presented and completely missed by the legacy one. We have also shown that there can be important cross terms of the phase and amplitude components.

For backscattering noise, we have retained the full nonlinear dependence on the backscatterer motion through the sine and cosine quadratures of the accumulated optical phase. This treatment distinguishes the small motion regime, where the scattered field is almost purely phase-like and does not generate appreciable AC radiation-pressure noise, from the large motion regime, where phase wrapping populates both quadratures and radiation pressure coupling becomes important. In this sense, the full model clarifies when the radiation pressure enhancement used in previous analytical estimates is physically justified and when it can lead to an overestimate of the scattered-light contribution. Additionally, we have shown that the cross power spectral density between the sine and cosine quadratures can enter the noise, an effect that had been ignored before.

We have also shown that the scattered-light transfer factors depend sensitively on the interferometer configuration. Variations of the homodyne readout angle, the signal-extraction cavity detuning, and the microscopic arm detuning can introduce optical-spring and anti-spring features, shift resonances, and modify the balance between amplitude and phase coupling. These effects are especially important for detector designs that employ a detuned signal extraction. Therefore, scattered-light requirements for next-generation observatories should not be evaluated independently of the optical operating point assumed for the detector.

Finally, we applied the formalism to the ET-LF configuration previously studied with legacy scattered-light models. The revised estimate shows that backscattering noise can be larger at low frequencies, mainly because of the detuned signal-extraction response, while diffraction noise is also modified once the correct radiation-pressure coupling and the linearity of the diffraction perturbation are taken into account. These differences demonstrate that accurate transfer factors are not merely a formal refinement, but can lead to non-negligible changes in the scattered-light noise budget of next-generation detectors.

Overall, the framework presented here provides a more complete and configuration-dependent mapping between optical scattering simulations and equivalent strain noise. It can be directly combined with FFT-based propagation tools such as SIS, measured scatterer vibration spectra, and realistic interferometer optical designs. This makes it a useful ingredient for the design of baffles, the evaluation of scattered-light requirements, and the interpretation of stray light couplings during commissioning and operation of current and future gravitational-wave observatories.

\begin{acknowledgments}
The author would like to thank K.~Kuns for continuous discussions and feedback on the manuscript. The author also thanks the feedback received by E. Vallejo-Pagès. This research has been supported by the National Science Foundation (NSF) awards PHY-2309064, PHY-2308793 and PHY-2308794. This document has received a LIGO DCC number of LIGO-P2600384.
\end{acknowledgments}

\bibliography{ref}

\appendix
\section{Phase-wrapping for diffraction}\label{app:phase_wrapping_diffraction}
Let's consider that the exact coupling of the field with the resonating TEM$_{00}$ mode is a non-linear function of the baffle motion, $x(t)$, such that
\begin{equation}
    a_{00}(x(t))=F(x(t))\, \in\mathbb{C}\,,
\end{equation}

Then, the extra coupling due to the baffle motion is 
\begin{equation}
    \Delta a_{00}(t) = F(x(t))-F(0)\, ,
\end{equation}
which can be taylor expanded for small displacements as 
\begin{equation}\label{eq:Delta_a_Taylor}
    \Delta a_{00}(t) = F'(0)x(t)+\frac{1}{2}F''(0)x^2(t)+\mathcal{O}(x^3(t))\, .
\end{equation}

In SIS, the quantity $\delta a_{00}$ is computed as~\cite{Romero21} 
\begin{equation}
    \delta a_{00} = \frac{a_{00}(+\delta)-a_{00}(-\delta)}{2\delta}\, ,
\end{equation}
which corresponds to the second order, central difference, numerical estimation of the first derivative around the undisplaced point. Therefore, $\delta a_{00}$ is equivalent to $F'(0)$ of Eq.~\eqref{eq:Delta_a_Taylor}, implying that we compute the diffraction coupling as a first order estimation:
\begin{equation}
    \Delta a_{00}(t)\approx F'(0)x(t)\, ,
\end{equation}
and in the frequency domain is equivalent to 
\begin{equation}
    \Delta a_{00}(f)\approx F'(0)\tilde{X}(f)\, ,
\end{equation}

This shows that the treatment done in SIS assumes the coupling is linear in the displacement and, therefore, one does not need to upconvert the noise in this regime. Nonetheless, one must take this effect into account when the truncation of the Taylor expansion of Eq.~\eqref{eq:Delta_a_Taylor} is at higher orders. For example, if instead of keeping only the linear term we would keep the second derivative as well then, in the frequency domains we get
\begin{equation}
    \Delta a_{00}(f) = F'(0)\tilde{X}(f)+\frac{1}{2}F''(0)(\tilde{X}*\tilde{X})(f)\, ,
\end{equation}
which explicitly shows that the second order term is not linear in the noise spectrum but rather with each convolution. This implies that if the second derivative is not negligible, one should, in theory, account for this nonlinear effect, that is equivalent to the phase wrapping in the backscattering case.

A way to see this effect in practice is through a small example. Consider a simple baffle movement of the form $x(t) = A \cos(\omega t)$. Then, by means of Eq.~\eqref{eq:Delta_a_Taylor} one obtains 
\begin{align*}
        \Delta a_{00}(t) &= F'(0)A\cos(\omega t) + \frac{1}{2}F''(0)A^2\cos^2(\omega t) \\ &= F'(0)A\cos(\omega t)+\frac{1}{4}F''(0)A^2(1+\cos(2\omega t))\, ,
\end{align*}
which shows that a second harmonic appears at $2\omega$, adding non-linearity to the final result.

A practical way to check if this is relevant is by computing the quotient between the two coefficients as 
\begin{equation}\label{eq:eta_nl}
    \eta_{\rm nl} \equiv \frac{|F''(0)|\sigma_x}{2|F'(0)|}\, ,
\end{equation}
where $\sigma_x$ is a representative RMS motion of the baffle. In general, we can split the motion into a slow and fast moving part as $x(t) = x_{\rm slow}(t)+x_{\rm fast}(t)$, where the slow moving part is at $f\ll 10$~Hz and the fast one around $f\gtrsim 10$~Hz~\cite{Flanagan94}. Assuming that the rms of the fast part is small (as it happens in realistic noise spectra), then we can approximate
$$
F(x_{\rm slow}+x_{\rm fast})\approx F(x_{\rm slow})+F'(x_{\rm slow})x_{\rm fast}\, ,
$$
which shows that while we care about the band of the fast vibrations, the coupling derivatives are mostly driven by those at small frequencies. Therefore, the RMS motion that is relevant to us is 
\begin{equation}
    \sigma_x = \sqrt{\int_{f_{\min}}^{f_{\rm c}}\tilde{X}^2(f) ~\td f}\, ,
\end{equation}
where we take $f_{\rm c}=10$~Hz and $f_{\min}$ dictated by the data available (for example, can be $f_{\min}=1/T$, where $T$ is the total observing time of the data). The final piece to evaluate Eq.~\eqref{eq:eta_nl} is the second derivative, but this can be estimated from SIS' outputs. With all of these, if we evaluate Eq.~\eqref{eq:eta_nl}, we can distinguish two regimes: $\eta_{\rm nl}\ll 1$, for which the linear approximation is fine; and $\eta_{\rm nl}\gtrsim1$, for which nonlinearities are relevant and one must account for them. 

\section{Derivation of the transfer factors from the node matrix formalism}\label{app:derivation}

In this appendix we rederive the optical transfer factors directly from the linear system associated with the signal-flow diagram. This is a standard derivation only reproduced here for completeness.

We collect the node fields in the ordered vector
\begin{equation}
\mathbf{x} \equiv
\begin{pmatrix}
\avec{sai}\\
\avec{sho}\\
\avec{iai}\\
\avec{iho}\\
\avec{ehi}\\
\avec{eho}\\
\avec{ihi}\\
\avec{iao}\\
\avec{shi}\\
\avec{sao}
\end{pmatrix},
\end{equation}
so that the optical relations can be written as
\begin{equation}
\mathbf{A}\,\mathbf{x} = \mathbf{b}.
\end{equation}
Here $\mathbf{A}$ is the block matrix
\begin{widetext}
\begin{equation}
\mathbf{A} =
\begin{pmatrix}
\mathbb{I} & 0 & 0 & 0 & 0 & 0 & 0 & 0 & 0 & 0 \\
-t_s\mathbb{I} & \mathbb{I} & 0 & 0 & 0 & 0 & 0 & 0 & r_s\mathbb{I} & 0 \\
0 & -\mathbb{L}_s & \mathbb{I} & 0 & 0 & 0 & 0 & 0 & 0 & 0 \\
0 & 0 & -t_i\mathbb{I} & \mathbb{I} & 0 & 0 & r_i\mathbb{I} & 0 & 0 & 0 \\
0 & 0 & 0 & -\mathbb{L}_a & \mathbb{I} & 0 & 0 & 0 & 0 & 0 \\
0 & 0 & 0 & 0 & -\mathbb{R}_e & \mathbb{I} & 0 & 0 & 0 & 0 \\
0 & 0 & 0 & 0 & 0 & -\mathbb{L}_a & \mathbb{I} & 0 & 0 & 0 \\
0 & 0 & -r_i\mathbb{I} & 0 & 0 & 0 & -t_i\mathbb{I} & \mathbb{I} & 0 & 0 \\
0 & 0 & 0 & 0 & 0 & 0 & 0 & -\mathbb{L}_s & \mathbb{I} & 0 \\
-r_s\mathbb{I} & 0 & 0 & 0 & 0 & 0 & 0 & 0 & -t_s\mathbb{I} & \mathbb{I}
\end{pmatrix}\, .
\label{eq:node_matrix_appendix}
\end{equation}
\end{widetext}

The inhomogeneous term $\mathbf{b}$ encodes the additional scattered-light field. These appear on the right hand side as we treat them as perturbations on top of the standard field. For the case of backscattering and diffraction here considered the perturbation appear in different nodes, so that the right hand side can be
\begin{equation}
\mathbf{b}_{\rm bs} =
\begin{pmatrix}
0\\
0\\
0\\
\sbs\\
0\\
0\\
0\\
0\\
0\\
0
\end{pmatrix},
\qquad
\mathbf{b}_{\rm diff} =
\begin{pmatrix}
0\\
0\\
0\\
0\\
\sdiff\\
0\\
0\\
0\\
0\\
0
\end{pmatrix},
\end{equation}

This corresponds to the backscattering perturbation $\sbs$ being injected in the $\avec{iho}$ equation, and the diffraction one $\sdiff$ in the $\avec{ehi}$ equation.

The matrix of Eq.~\eqref{eq:node_matrix_appendix} contains two independent feedback cycles: the arm-cavity loop ($\avec{iho}\rightarrow \avec{ehi}\rightarrow \avec{eho}\rightarrow \avec{ihi}\rightarrow \avec{iho}$),
and the signal-extraction loop ($\avec{shi}\rightarrow \avec{sho}\rightarrow \avec{iai}\rightarrow \avec{iao}\rightarrow \avec{shi}$).

Therefore, a convenient choice of loop-closing variables is 
\begin{equation}
\mathbf{z}\equiv
\begin{pmatrix}
\avec{iho}\\
\avec{shi}
\end{pmatrix}.
\end{equation}

Once $\mathbf{z}$ is fixed, all the remaining fields can be obtained by forward substitution as there is no further recursion appearing in those equations. We thus define
\begin{equation}
\mathbf{y}\equiv
\begin{pmatrix}
\avec{sai}\\
\avec{sho}\\
\avec{iai}\\
\avec{ehi}\\
\avec{eho}\\
\avec{ihi}\\
\avec{iao}\\
\avec{sao}
\end{pmatrix},
\qquad
\mathbf{x}=
\begin{pmatrix}
\mathbf{y}\\
\mathbf{z}
\end{pmatrix},
\end{equation}
and partition the linear system as
\begin{equation}
\begin{pmatrix}
\mathbf{A}_{yy} & \mathbf{A}_{yz}\\
\mathbf{A}_{zy} & \mathbf{A}_{zz}
\end{pmatrix}
\begin{pmatrix}
\mathbf{y}\\
\mathbf{z}
\end{pmatrix}
=
\begin{pmatrix}
\mathbf{b}_y\\
\mathbf{b}_z
\end{pmatrix}.
\end{equation}

From the first block row,
\begin{equation}
\mathbf{A}_{yy}\mathbf{y}+\mathbf{A}_{yz}\mathbf{z}=\mathbf{b}_y,
\end{equation}
so that
\begin{equation}
\mathbf{y}=-\mathbf{A}_{yy}^{-1}\mathbf{A}_{yz}\mathbf{z}+\mathbf{A}_{yy}^{-1}\mathbf{b}_y.
\end{equation}
Substituting this into the second block row gives the Schur complement system
\begin{equation}
\mathbf{S}\,\mathbf{z}=\mathbf{b}_z-\mathbf{A}_{zy}\mathbf{A}_{yy}^{-1}\mathbf{b}_y,
\qquad
\mathbf{S} \equiv \mathbf{A}_{zz}-\mathbf{A}_{zy}\mathbf{A}_{yy}^{-1}\mathbf{A}_{yz}.
\label{eq:schur_general}
\end{equation}

For the present matrix, carrying out the forward substitution yields
\begin{equation}
\mathbf{S}=
\begin{pmatrix}
\mathbb{I}+r_i \mathbb{E}_a & t_i \mathbb{L}_s r_s\\
-t_i\mathbb{L}_s  \mathbb{E}_a & \mathbb{I}+ r_ir_s\mathbb{L}_s \mathbb{L}_s 
\end{pmatrix},
\qquad
\mathbb{E}_a \equiv \mathbb{L}_a \mathbb{R}_e \mathbb{L}_a.
\label{eq:reduced_loop_matrix}
\end{equation}

For backscattering, the source enters directly in the first component of $\mathbf{z}$, so Eq.~\eqref{eq:schur_general} becomes
\begin{equation}
\mathbf{S}
\begin{pmatrix}
\avec{iho}\\
\avec{shi}
\end{pmatrix}
=
\begin{pmatrix}
\sbs\\
0
\end{pmatrix}.
\label{eq:reduced_bs}
\end{equation}

For diffraction, the perturbation enters in the eliminated sector $\mathbf{y}$ and therefore induces an effective source in the reduced loop system. One finds
\begin{equation}
\mathbf{S}
\begin{pmatrix}
\avec{iho}\\
\avec{shi}
\end{pmatrix}
=
\begin{pmatrix}
-r_i \mathbb{L}_a \mathbb{R}_e\\
t_i \mathbb{L}_s  \mathbb{L}_a \mathbb{R}_e
\end{pmatrix}
\sdiff.
\label{eq:reduced_diff}
\end{equation}

The readout field is obtained from the last row of the eliminated sector, namely
\begin{equation}
\avec{sao}=t_s\,\avec{shi}.
\label{eq:output_relation_appendix}
\end{equation}

To solve these two equations we write Eq.~\eqref{eq:reduced_loop_matrix} as
\begin{equation}
\mathbf{S}=
\begin{pmatrix}
\mathbf{S}_{11} & \mathbf{S}_{12}\\
\mathbf{S}_{21} & \mathbf{S}_{22}
\end{pmatrix},
\end{equation}
because we have that for a generic system
\begin{equation}
\begin{pmatrix}
\mathbf{S}_{11} & \mathbf{S}_{12}\\
\mathbf{S}_{21} & \mathbf{S}_{22}
\end{pmatrix}
\begin{pmatrix}
\mathbf{x}\\
\mathbf{y}
\end{pmatrix}
=
\begin{pmatrix}
\mathbf{f}\\
\mathbf{g}
\end{pmatrix},
\end{equation}
the second component is
\begin{equation}
\mathbf{y}=(\mathbf{S}_{22}-\mathbf{S}_{21}\mathbf{S}_{11}^{-1}\mathbf{S}_{12})^{-1}(\mathbf{g}-\mathbf{S}_{21}\mathbf{S}_{11}^{-1}\mathbf{f}).
\label{eq:block_solution_y}
\end{equation}
Applying Eq.~\eqref{eq:block_solution_y} to Eqs.~\eqref{eq:reduced_bs}--\eqref{eq:reduced_diff} gives the field $\avec{shi}$ directly. It is convenient to reduce the notation  by defining the effective arm reflectivity as
\begin{equation}
\mathbb{R}_a \equiv r_i + t_i^2 (\mathbb{I}+r_i \mathbb{E}_a)^{-1}\mathbb{E}_a.
\label{eq:Ra_appendix}
\end{equation}
Using this definition, the Schur complement of the upper-left block of $\mathbf{S}$, which is the term inside the first parenthesis in Eq.~\eqref{eq:block_solution_y}, becomes
\begin{equation}
\mathbf{S}_{22}-\mathbf{S}_{21}\mathbf{S}_{11}^{-1}\mathbf{S}_{12} = 
\mathbb{I}+r_s \mathbb{L}_s \mathbb{R}_a \mathbb{L}_s.
\label{eq:sec_dressed_denom}
\end{equation}

For Eq.~\eqref{eq:reduced_bs}, one has $\mathbf{f}=\sbs$ and $\mathbf{g}=0$. Using Eq.~\eqref{eq:block_solution_y},
\begin{equation}
\avec{shi}^{(\rm bs)}
=
t_i\left[\mathbb{I}+r_s \mathbb{L}_s \mathbb{R}_a \mathbb{L}_s\right]^{-1}
\mathbb{L}_s  (\mathbb{I}+r_i \mathbb{E}_a)^{-1} \mathbb{E}_a\,\sbs.
\end{equation}
Then, from Eq.~\eqref{eq:output_relation_appendix},
\begin{equation}
\avec{sao}^{(\rm bs)}
=
t_st_i\left[\mathbb{I}+r_s \mathbb{L}_s \mathbb{R}_a \mathbb{L}_s\right]^{-1}
\mathbb{L}_s  (\mathbb{I}+r_i \mathbb{E}_a)^{-1} \mathbb{E}_a\,\sbs,
\label{eq:appendix_bs_transfer}
\end{equation}
which gives the appropriate transfer factor for backscattering.

On the other hand, for Eq.~\eqref{eq:reduced_diff}, one has
\begin{equation}
\mathbf{f}=-r_i \mathbb{L}_a \mathbb{R}_e\,\sdiff,
\qquad
\mathbf{g}=t_i \mathbb{L}_s \mathbb{L}_a \mathbb{R}_e\,\sdiff.
\end{equation}

Using Eq.~\eqref{eq:block_solution_y} and Eq.~\eqref{eq:output_relation_appendix}, we get the relation of the diffraction perturbation to the field at end of the signal extraction cavity as 
\begin{equation}
\avec{shi}^{(\rm diff)}
=t_st_i\left[\mathbb{I}+r_s \mathbb{L}_s \mathbb{R}_a \mathbb{L}_s\right]^{-1}
\mathbb{L}_s  (\mathbb{I}+r_i \mathbb{E}_a)^{-1} \mathbb{L}_a\mathbb{R}_e\,\sdiff,
\end{equation}
where the relation 
\begin{equation}
\mathbb{I}-\mathbb{E}_a(\mathbb{I}+r_i \mathbb{E}_a)^{-1}r_i=(\mathbb{I}+r_i \mathbb{E}_a)^{-1},
\end{equation}
has been used. This relation follows from the fact that $(\mathbb{I}+r_i \mathbb{E}_a)^{-1}$ is a function of $\mathbb{E}_a$ and therefore commutes with it.

Finally, one must project the outgoing SEM field onto the readout quadrature multiplying these outputs by $\vvec^\dagger$ (see Eq.~\ref{eq:readout}). Therefore, the two transfer factors become
\begin{equation}
    \Tvec{i}^\dagger=\frac{t_st_i}{\sqrt{2}}\vvec^\dagger\left[\mathbb{I}+r_s \mathbb{L}_s \mathbb{R}_a \mathbb{L}_s\right]^{-1}
\mathbb{L}_s  (\mathbb{I}+r_i \mathbb{E}_a)^{-1} \mathbb{E}_a\, ,
\end{equation}
and 
\begin{equation}
    \Tvec{e}^\dagger=\frac{t_st_i}{\sqrt{2}}\vvec^\dagger\left[\mathbb{I}+r_s \mathbb{L}_s \mathbb{R}_a \mathbb{L}_s\right]^{-1}
\mathbb{L}_s  (\mathbb{I}+r_i \mathbb{E}_a)^{-1} \mathbb{L}_a\mathbb{R}_e\,,
\end{equation}
where the factor $1/\sqrt{2}$ has been added to account for the effect of the beam splitter~\cite{Hall:2024vsc}. These are the desired transfer factors derived directly from the node matrix formalism.

\section{Equivalence of diffraction source node}
\label{app:diffraction_node}

In the main text, the diffraction perturbation computed by a stationary FFT simulations such as SIS is injected into the incoming HR side of the ETM, i.e. at the \(\avec{ehi}\) node. This convention allows the perturbation to undergo the ETM optomechanical reflection operator \(\mathbb{R}_e\), thereby accounting for the radiation-pressure associated with the amplitude quadrature of the clipped field. In this appendix we justify why, despite the fact that SIS computes the fractional diffraction perturbation after the ETM reflection, we can consider the perturbation as incoming and not outgoing.

Let \(\psi_{00}(\vec r)\) denote the unperturbed incident fundamental mode on the ETM, with
\begin{equation}
P \equiv \langle \psi_{00}(\vec r),\psi_{00}(\vec r)\rangle = \int_A |\psi_{00}(\vec r)|^2\,\td A .
\end{equation}
The baffle-induced field perturbation incident on the ETM can be decomposed as
\begin{equation}
\delta \psi_{\rm in}(\vec r)
=
\delta a_{\rm in}\,\psi_{00}(\vec r)
+
\delta \psi_{\perp}(\vec r),
\label{eq:incident_diff_decomposition}
\end{equation}
where \(\delta a_{\rm in}\) is the true incident TEM$_{00}$ fractional perturbation and
\(\delta \psi_{\perp}\) contains the part orthogonal to the carrier mode,
\begin{equation}
\int_a \psi_{00}^*(\vec r)\,\delta \psi_{\perp}(\vec r)\,\td A = 0 .
\end{equation}

At linear order, the radiation-pressure force is driven by the perturbation of the carrier. Therefore, in the absence of additional mode mixing, only the TEM$_{00}$ fractional component $\delta a_{\rm in}$ contributes to the DARM radiation-pressure response. The orthogonal component, $\delta \psi_{\perp}$, can contribute only through higher-order terms and mode mixing. We aim at bounding the mode-mixing contribution produced by the ETM phase map.

The static ETM reflection operator used by the FFT simulation can be written as~\cite{Romero21}
\begin{equation}
\widehat{R}_{\rm mirr}
=
-r_e\exp\left[i\phi(\vec r)\right],
\qquad
\phi(\vec r)=\frac{4\pi h(\vec r)}{\lambda},
\label{eq:static_map_operator}
\end{equation}
where \(h(\vec r)\) is the ETM height map after removing piston, pitch, and yaw. SIS returns the fractional TEM$_{00}$ perturbation after this reflection. In the notation above,
this quantity can be written as
\begin{equation}
\delta a_{\rm out}
=
\frac{
\displaystyle\int_A \psi_{00}^*(\vec r)\,\widehat{R}_{\rm mirr}\,
\delta \psi_{\rm in}(\vec r)\,\td A
}{
\displaystyle\int_A \psi_{00}^*(\vec r)\,\widehat{R}_{\rm mirr}\,
\psi_{00}(\vec r)\,\td A
}.
\label{eq:delta_a_sis_definition}
\end{equation}

If \(\widehat{R}_{\rm mirr}\) were a scalar operator, the numerator and denominator would be multiplied by the same factor and one would have exactly \(\delta a_{\rm out}=\delta a_{\rm in}\). The only possible difference comes from the non-scalar part of the phase-map operator, which can scatter the orthogonal diffracted field \(\delta \psi_{\perp}\) back into the TEM$_{00}$ projection.

For small surface roughness, which is typical in GW detector optics, \(|\phi|\ll1\), and we can expand
\begin{equation}
\widehat{R}_{\rm mirr}
\simeq
-r_e\left[1+i\phi(\vec r)\right].
\end{equation}
The scalar factor \(-r_e\) cancels in Eq.~\eqref{eq:delta_a_sis_definition}. Substituting
Eq.~\eqref{eq:incident_diff_decomposition} and expanding to first order in \(\phi\), we find
\begin{equation}
\delta a_{\rm out}
\simeq
\delta a_{\rm in}
+
i\left[
\frac{1}{P}
\int_A
\psi_{00}^*(\vec r)\,
\phi(\vec r)\,
\delta \psi_{\perp}(\vec r)\,
\td A
\right].
\label{eq:delta_a_sis_expanded}
\end{equation}
This shows that the residual error is
\begin{equation}
\xi_{\rm map}
\equiv
\delta a_{\rm out}-\delta a_{\rm in}
\simeq
\frac{i}{P}
\int_A
\psi_{00}^*(\vec r)\,
\phi(\vec r)\,
\delta \psi_{\perp}(\vec r)\,
\td A .
\label{eq:xi_map_definition}
\end{equation}
Equivalently, we can subtract the beam-weighted piston phase
\begin{equation}
\bar{\phi}
=
\frac{1}{P}
\int_A |\psi_{00}(\vec r)|^2\phi(\vec r)\,\td A ,
\end{equation}
so that the residual error is computed as
\begin{equation}
\xi_{\rm map}
\simeq
\frac{i}{P}
\int_A
\psi_{00}^*(\vec r)\,
\left[\phi(\vec r)-\bar{\phi}\right]\,
\delta \psi_{\perp}(\vec r)\,
\td A ,
\label{eq:xi_map_piston_subtracted}
\end{equation}
which makes explicit that only the nontrivial spatial structure of the map produces mode
mixing.

Applying the Cauchy--Schwarz inequality to Eq.~\eqref{eq:xi_map_piston_subtracted} gives
the absolute bound
\begin{multline}
|\xi_{\rm map}|
\le
\frac{1}{P}
\left[
\int_A
|\psi_{00}(\vec r)|^2
|\phi(\vec r)-\bar{\phi}|^2
\td A
\right]^{1/2}\\ \times
\left[
\int_A
|\delta \psi_{\perp}(\vec r)|^2
\td A
\right]^{1/2}.
\end{multline}
Defining the beam-weighted RMS height of the ETM map as
\begin{equation}
\sigma_h^2
\equiv
\frac{1}{P}
\int_A
|\psi_{00}(\vec r)|^2
\left[h(\vec r)-\bar h\right]^2
\td A ,
\end{equation}
and the incident power in HOMs as
\begin{equation}
P_{\perp}
\equiv
\int_A
|\delta \psi_{\perp}(\vec r)|^2
\td A ,
\end{equation}
we obtain
\begin{equation}
|\xi_{\rm map}|
\le
\frac{4\pi\sigma_h}{\lambda}
\sqrt{\frac{P_{\perp}}{P}}\, .
\label{eq:xi_map_bound}
\end{equation}
For $\lambda=1064\,{\rm nm}$, we have
\begin{equation}
\frac{4\pi\sigma_h}{\lambda}
\simeq
1.18\times10^{-2}
\left(
\frac{\sigma_h}{1\,{\rm nm}}
\right)\, .
\end{equation}

The error is further suppressed by the fractional power in HOMs relative to the circulating power, which for realistic conditions is $\sqrt{P_\perp/P}\ll 1$. Therefore, we can assume that the couplings before and after reflection are equivalent, $\delta a_{\rm out}\simeq \delta a_{\rm in}$, with an error bounded by Eq.~\eqref{eq:xi_map_bound}.

\end{document}